\documentclass[preprint,aps,prc,superscriptaddress,showpacs,floatfix]{revtex4-1}
\usepackage{graphicx}
\usepackage{wrapfig}
\usepackage{dcolumn}
\usepackage{bm}
\usepackage{amsfonts}
\usepackage{amsmath}
\usepackage{cancel}
\usepackage{multirow}
\usepackage{color}
\usepackage{ulem}
\begin{document}
\title{Non-symmetrized hyperspherical harmonic basis for $A$--bodies}

\author{M. Gattobigio}
\affiliation{Universit\'e de Nice-Sophia Antipolis, Institut Non-Lin\'eaire de
Nice,  CNRS, 
1361 route des Lucioles, 06560 Valbonne, France }
\author{A. Kievsky}
\author{M. Viviani}
\affiliation{Istituto Nazionale di Fisica Nucleare, Largo Pontecorvo 3, 56100 Pisa, Italy}

\begin{abstract}
The use of the hyperspherical harmonic (HH) basis in the description of
bound states in an $A$-body system composed by identical particles
is normally preceded by a symmetrization procedure in which the statistic 
of the system is taken into account. This 
preliminary step is not strictly necessary; 
the direct use of the HH basis is possible, even if the basis has not
a well defined behavior under particle permutations.
In fact, after the diagonalization of the Hamiltonian matrix, the 
eigenvectors reflect the symmetries present in it. 
They have well defined symmetry under particle permutation and the
identification of the physical states is possible, as it will be shown in
specific cases. 
The problem related to the large degeneration of the 
basis is circumvented by constructing the 
Hamiltonian matrix as a 
sum of products of sparse matrices. This particular
representation of the 
Hamiltonian is well suited for a numerical 
iterative diagonalization, where only the action of the matrix 
on a vector is needed.
As an example we compute bound states for systems with
$A=3-6$ particles interacting through
a short-range central interaction. We also consider the case in which the
potential is restricted to act in relative $s$-waves with and without
the inclusion of the Coulomb potential. This very simple model predicts
results in qualitative good agreement with the experimental data and
it represents a first step in a project dedicated to the use of the HH basis to
describe bound and low energy scattering states in light nuclei.
\end{abstract}

\maketitle

\section{Introduction}

The {\sl ab initio} description of light nuclear systems, starting from the
nucleon-nucleon (NN) interaction, requires well established methods to solve
the Schr\"odinger equation. Among them, the Green function Monte Carlo (GFMC) 
method has been extensively used to describe light nuclei up to $A=10$ and the
no-core shell model (NCSM) up to $A=12$~\cite{gfmc,ncsm}. In the
$A\le4$ systems, well established methods for treating 
both bound and scattering states exist as the
Faddeev equations ($A=3$) and
the Faddeev-Yakubovsky equations ($A=4$) in configuration or momentum space, 
and the Hyperspherical Harmonic (HH) expansion. 
All these
methods have proven to be of great accuracy and they have been tested using
different benchmarks~\cite{benchmark1,benchmark2,benchmark3}. 

The HH method provides a systematic way of constructing a 
complete basis for the
expansion of the $A$-particle wave function
and its use in the $A>4$ systems has been subject 
of intense investigations over the last years. In the specific case of
application to nuclear physics, the wave function has to be antisymmetric and,
therefore, the HH basis has been managed to produce basis states having
well defined properties under
particle permutations. Different schemes to construct hyperspherical functions 
with an arbitrary permutational symmetry are given in 
Refs.~\cite{novo94,barnea95,barnea99}. Recently, a procedure
for constructing HH functions in terms of a single particle basis has
been proposed in Ref.~\cite{timofeyuk:08}.

In a different approach, the authors have used the HH basis, without a previous
symmetrization procedure, to describe bound states in three- and four-particle
systems~\cite{mario}. 
It has been observed that the eigenvectors of the Hamiltonian matrix
reflects the symmetries present in it, even if it has been constructed using
the non-symmetrized basis. The only requirement was to include all the HH basis
elements having the same grand angular quantum number $K$. It is a 
property of the HH basis that basis elements having well defined behavior
under particle permutation can be constructed as a linear combination of
HH elements having the same value of $K$. Therefore, if 
the Hamiltonian commutes with the group of permutations of $A$ objects, $S_A$, 
the diagonalization procedure generates eigenvectors having
well defined permutation symmetry that can be organized in accordance
with the irreducible representations of $S_A$.
Moreover, identifying those eigenvectors with the desired
symmetry, the corresponding energies can be considered variational estimates.
In particular, in Ref.~\cite{mario}, it was possible to identify a
subset of eigenvectors and eigenvalues corresponding exactly to those that
would be obtained performing the preliminary symmetrization of the states.
It should be noticed that
the simplicity of using the HH basis without a preliminary antisymmetrization
step, has to be counterbalanced with the large dimension of the matrices to
be diagonalized. However, at present, different techniques are available to
treat (at least partially) this problem.

In the present article we 
continue the study of the non-symmetrized HH basis, extending the applications to
systems with $A>4$. 
In pursuit of this goal, we have developed a particular representation
of the Hamiltonian matrix, which is systematic with respect to
the number of particles and well suited for a numerical implementation.
As mentioned, one of the main problem in using the HH basis is its large
degeneracy, resulting in very large matrices. On the other hand, the potential
energy matrix, expressed as a sum of pairwise interactions,
cannot connect arbitrary basis elements
differing in some specific quantum numbers. This means that in
some representation each pairwise-interaction term has to be represented by a
sparse matrix.  For example, the matrix representation of the
potential $V(1,2)$, constructed in terms of basis
elements in which the quantum numbers of particles $(1,2)$ are well
defined, is sparse in $A\ge3$ systems. In fact, its matrix elements
connecting basis elements with different quantum numbers labelling states which
do not
involve particles $(1,2)$ are zero. A problem arises when the matrix elements of 
the generic term $V(i,j)$, defining the interaction between particles $(i,j)$, has 
to be calculated using basis elements in which the quantum numbers of
particles $(i,j)$ are not well defined. One operative way to solve this
problem consists in 
rotating the basis to a system of coordinates in which 
particles $(i,j)$ have well defined quantum numbers. This makes the matrix $V(i,j)$ 
sparse. However, we would like the rotation matrix to be sparse too,
which in general
it is not true. This last problem is solved noticing that the rotation
matrix can be expressed as a product of sparse matrices, each one representing
a rotation which involves a permutation of particles of successive numbering. After 
these manipulations the potential energy matrix results in a sum of products of 
sparse matrices suitable for numerical implementations.

An advantage in using the non-symmetrized HH basis appears when symmetry
breaking terms are present in the Hamiltonian. In the case of the nuclear
Hamiltonian with charge-symmetry breaking terms, this means that different
total isospin components $T$ are present in the wave function. 
For example, the three-nucleon bound state wave function includes $T=1/2,3/2$
components and the four-nucleon bound state wave function includes $T=0,1,2$
components, requiring the inclusion of different spatial symmetries in the
wave function. Therefore,
considering all the possible spin and isospin components, the number of HH
states having well defined spatial symmetries, necessary to construct the wave 
function, and the dimension of the non-symmetrized basis is comparable. 
High isospin
components are in general a small part of the total wave function. They
are difficult to include in the antisymmetrized basis since appreciably
increases the number of basis elements and, at the same time, they improve
very little the description of the state.
In practical cases they are disregarded, or
partially included, with the consequence that the occupation 
probabilities of the high isospin states are not always well determined 
(see Ref.~\cite{viviani05}). Conversely, using the non-symmetrized basis, all the 
isospin components are automatically generated. As an example we will show results 
for $A=3-6$ systems using short-range central interactions with and without the 
inclusion of the Coulomb potential. 

To summarize, in this paper we present the implementation
of the non-symmetrized HH basis for $A$-body system using the factorization
of the potential energy matrix mentioned before. In order to give a detailed description
of this construction, we consider only spatial degrees of freedom;
accordingly, we show examples using a central interaction. The diagonalization
of the Hamiltonian produces eigenvectors organized in multiplets of the
dimension of the corresponding irreducible representation of $S_A$, and the
different symmetries will be identified using the appropriate Casimir operator.
Not all the states belonging to a particular representation can be antisymmetrized
using the spin-isospin functions of $A$ nucleons and, therefore, these states
are not physical. It should be noticed that the physical states could appear
in very high positions of the spectrum, in particular this is the case
for $A>4$ systems. On the other hand, the iterative methods, as the Lanczos method,
used to search selected eigenvalues and eigenvectors of large matrices are
more efficient for the extreme ones. To this respect, we have found
very convenient to use the symmetry-adapted Lanczos method
proposed in Ref.~\cite{slanczos},
which restricts the search to those states having a particular symmetry.
When possible, comparisons to 
different results in the literature will be done. Since we have in mind the 
description of light nuclear systems using realistic interactions, 
this study can be considered a preliminary step in the use of this technique. 

The paper is organized as follows; section \ref{sec:hh} is devoted to
a brief description of the HH basis. In sections \ref{sec:pot} the expression
for the potential energy matrix in terms of HH states is given. In section 
\ref{sec:results}
the results for the models proposed are shown. Section \ref{sec:conclu}
includes a brief discussion of the results and the perspectives of the
present work.

\section{The Harmonic Hyperspherical basis for $A$ bodies}\label{sec:hh}
In this section we introduce the notation and we present a brief overview of 
the properties of the HH basis.

\subsection{Basic properties of the HH basis}

In accord with Ref.\cite{mario},
we start with the following definition of the Jacobi
coordinates for an $A$ body system
with Cartesian coordinates $\mathbf r_1 \dots \mathbf r_A$

\begin{equation}
  \mathbf x_{N-j+1} = \sqrt{\frac{2 m_{j+1} M_j}{(m_{j+1}+M_j)m} } \,
                  (\mathbf r_{j+1} - \mathbf X_j)\,,
   \qquad
   j=1,\dots,N\,,
  \label{eq:jacobiCoordinates}
\end{equation}
where $m$ is a reference mass, $N=A-1$, and where we have defined
\begin{equation}
  M_j = \sum_{i=1}^j m_i\,, \qquad \mathbf X_j = \frac{1}{M_j} \sum_{i=1}^j
  m_i\mathbf r_i \,.
  \label{}
\end{equation}
Let us note that if all the masses are equal, $m_i = m\,$,
Eq.~(\ref{eq:jacobiCoordinates}) simplifies to
\begin{equation}
  \mathbf x_{N-j+1} = \sqrt{\frac{2 j}{j+1} } \,
                  (\mathbf r_{j+1} - \mathbf X_j)\,,
   \qquad
   j=1,\dots,N\,.
  \label{eq:jc2}
\end{equation}
For a given set of Jacobi coordinates $\mathbf x_1, \dots, \mathbf x_N$, 
we can introduce the hyperradius $\rho$
\begin{equation}
  \rho = \bigg(\sum_{i=1}^N x_i^2\bigg)^{1/2}
   = \bigg(2\sum_{i=1}^A (\mathbf r_i - \mathbf X)^2\bigg)^{1/2}
   = \bigg(\frac{2}{A}\sum_{j>i}^A (\mathbf r_j - \mathbf r_i)^2\bigg)^{1/2} \,,
  \label{}
\end{equation}
and the hyperangular coordinates $\Omega_N$
\begin{equation}
  \Omega_N = (\hat x_1, \dots, \hat x_N, \phi_2, \dots, \phi_N) \,,
  \label{}
\end{equation}
with the hyperangles $\phi_i$ defined via
\begin{equation}
  \cos\phi_i = \frac{x_i}{\sqrt{x_1^2 + \dots + x_i^2}}\,,\qquad i=2,\dots, N\,.
  \label{}
\end{equation}
The radial components of the Jacobi coordinates can be expressed in terms of
the hyperspherical coordinates
\begin{equation}
  \begin{aligned}
    &x_N = \rho \cos\phi_N \\
    &x_{N-1} = \rho \sin\phi_N \cos\phi_{N-1} \\
    &\qquad\vdots \\
    &x_{i} = \rho \sin\phi_N \cdots \sin\phi_{i+1}\cos\phi_i\\
    &\qquad\vdots \\
    &x_{2} = \rho \sin\phi_N \cdots \sin\phi_{3}\cos\phi_2  \\
    &x_{1} = \rho \sin\phi_N \cdots \sin\phi_{3}\sin\phi_2  \,. \\
  \end{aligned}
  \label{eq:hyp1} 
\end{equation}
Using the above hyperspherical angles $\Omega_N$, the surface
element becomes
\begin{equation}
  d\Omega_N = \sin\theta_1\,d\theta_1\,d\varphi_1
               \prod_{j=2}^N \sin\theta_j\,d\theta_j\,d\varphi_j
               (\cos\phi_j)^2 (\sin\phi_j)^{3j-4} d\phi_j \,,
  \label{eq:surface}
\end{equation}
and the Laplacian operator 
\begin{equation}
  \Delta = \sum_{i=1}^N \nabla_{\mathbf x_i}^2 = \left(
  \frac{\partial^2}{\partial\rho^2} +
  \frac{3N-1}{\rho}\frac{\partial}{\partial\rho} +
  \frac{\Lambda_N^2(\Omega_N)}{\rho^2}\right) \,,
  \label{}
\end{equation}
where the $\Lambda_N^2(\Omega_N)$ is the generalization of the angular momentum
and is called grand angular operator.

The HH functions ${\mathcal Y}_{[K]}(\Omega_N)$ are the eigenvectors 
of the grand angular momentum operator

\begin{equation}
  \bigg(\Lambda_N^2(\Omega_N) + K(K+3N-2)\bigg) {\mathcal Y}_{[K]}(\Omega_N) =
  0 \,.
  \label{}
\end{equation}
They can be expressed
in terms of the usual harmonic functions $Y_{lm}(\hat x)$ and of the
Jacobi polynomials $P_n^{a,b}(z)$. In fact, the explicit expression for the HH
functions is
\begin{equation}
  {\mathcal Y}_{[K]}(\Omega_N) = 
    \left[\prod_{j=1}^N Y_{l_jm_j}(\hat x_j) \right] \left[ \prod_{j=2}^N
    \,^{(j)}\!
{\mathcal P}_{K_j}^{\alpha_{l_j},\alpha_{K_{j-1}}}(\phi_j)\right] \,,
  \label{eq:hh}
\end{equation}
where $[K]$ stands for the set of quantum numbers $l_1,\dots,l_N,m_1, \dots,m_N,
n_2, \dots, n_N$, and the hyperspherical polynomial is
\begin{equation}
^{(j)}{\mathcal P}_{K_j}^{\alpha_{l_j},\alpha_{K_{j-1}}}(\phi_j) = 
{\mathcal
N}_{n_j}^{\alpha_{l_j},\alpha_{K_j}} 
(\cos\phi_j)^{l_j} (\sin\phi_j)^{K_{j-1}} 
P^{\alpha_{K_{j-1}}, \alpha_{l_j}}_{n_j}(\cos2\phi_j) \,,
\end{equation}
with the $K_j$ quantum numbers defined as
\begin{equation}
  K_j = \sum_{i=1}^j (l_i + 2n_i)\,, \qquad n_1 = 0\,, \qquad K \equiv K_N\, .
  \label{}
\end{equation}
The normalization factor is
\begin{equation}
  {\cal N}_{n}^{\alpha\beta} =
  \sqrt{\frac{2(2n+\alpha+\beta+1) n!\,
  \Gamma(n+\alpha+\beta+1)}{\Gamma(n+\alpha+1)\Gamma(n+\beta+1)}}\,,
  \label{eq:norma}
\end{equation}
where, for the special choice of hyperangles given by Eq.~(\ref{eq:hyp1}),
$\alpha_{K_j} = K_j +3j/2-1$ and $\alpha_{l_j}=l_j+1/2$.
The quantum number $K\equiv K_N$ is also known as the grand angular momentum.

The HH functions are normalized
\begin{equation}
  \int d\Omega_N \bigg({\mathcal Y}_{[K']}(\Omega_N)\bigg)^* {\mathcal Y}_{[K]}(\Omega_N)
   = \delta_{[K],[K']} \,,
  \label{eq:normalization}
\end{equation}
moreover, the HH basis is complete
\begin{equation}
  \sum_{[K]} \bigg({\mathcal Y}_{[K]}(\Omega_N)\bigg)^* {\mathcal
  Y}_{[K]}(\Omega'_N)  = \delta^{3N-1}(\Omega'_N - \Omega_N)\,.
  \label{}
\end{equation}

With the above definitions, the HH functions do not have well defined total
orbital angular momentum $L$ and $z$-projection $M$. It is possible to
construct HH functions having well defined values of $LM$ by coupling the
functions $Y_{l_jm_j}(\hat x_j)$. This can be achieved using different coupling
schemes. Accordingly, we can define the following HH function
\begin{equation}
    {\mathcal Y}^{LM}_{[K]}(\Omega_N) =
    \left[\prod_{j=2}^N \,
^{(j)}{\mathcal P}_{K_j}^{\alpha_{l_j},\alpha_{K_{j-1}}}(\phi_j)\right]
    \bigg[Y_{l_1}(\hat x_1) \otimes Y_{l_2}(\hat x_2)|_{L_2}
    \ldots \otimes Y_{l_{N-1}}(\hat x_{N-1})|_{L_{N-1}} 
    \otimes Y_{l_N}(\hat x_N) \bigg]_{LM}  \,,
\label{eq:hh0}
\end{equation}
having well defined values of $LM$, with the particular coupling scheme 
in which particles $(1,2)$ are coupled to $L_2$, which in turns, with $l_3$, 
is coupled to $L_3$ and so on, generating $N-2$ intermediate $L_i$-values.
The set of quantum numbers $[K]$ includes the $n_2\ldots n_N$ indices of
the Jacobi polynomials, the $l_1\ldots l_N$ angular momenta of the particles and
the intermediate couplings $L_2 \ldots L_{N-1}$. 

\begin{figure}
  \begin{center}
 \includegraphics[width=0.9\linewidth]{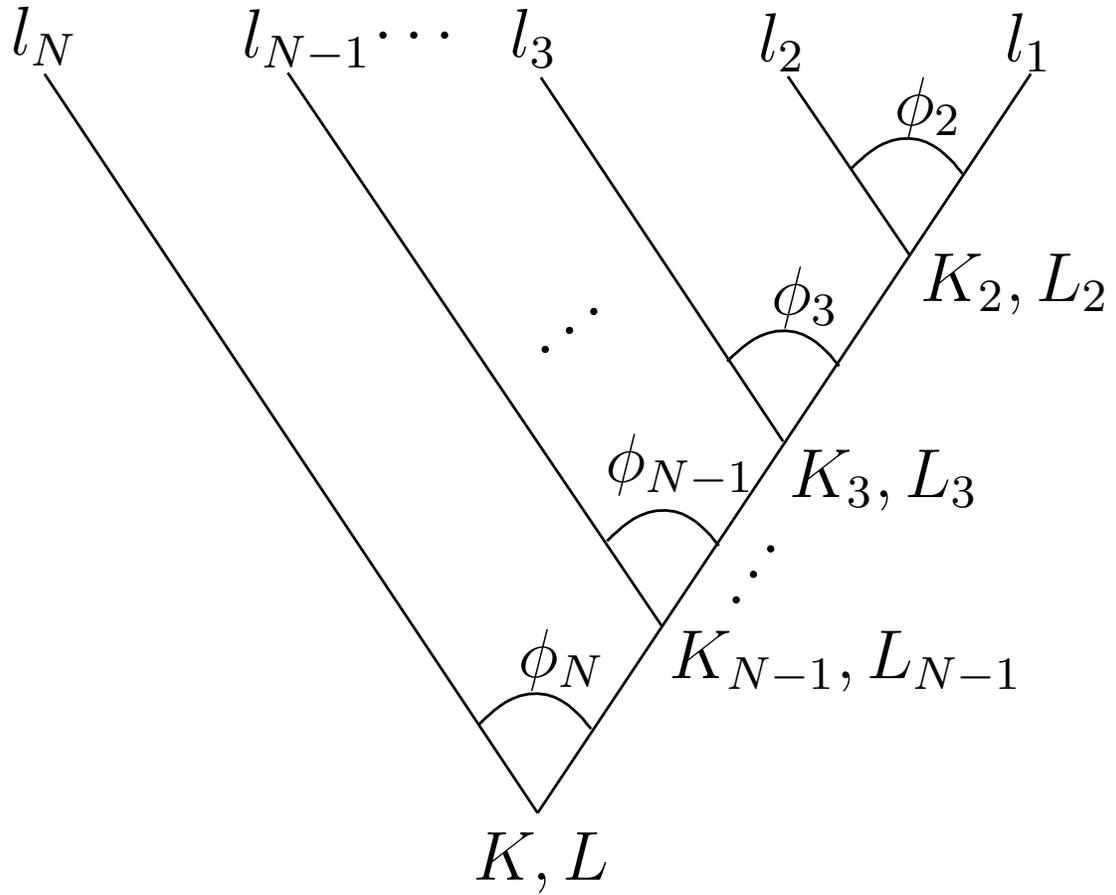}
  \end{center}
  \caption{Hyperspherical tree corresponding to 
Eq.({~\protect\ref{eq:hyp1}})}
 \label{fig:tree}
\end{figure}

In the definition of the hyperspherical coordinates in terms of
the radial components of the Jacobi coordinates it is useful to
introduce the hyperspherical tree structure~\cite{vilenkin}. For
example, the particular choice of Eq.~(\ref{eq:hyp1}), in the
coupling scheme of Eq.(~\ref{eq:hh0}),corresponds
to the one depicted in Fig.~\ref{fig:tree}, 
where we can also read the above-mentioned angular-momentum
coupling scheme. However, other definitions are
possible, and the corresponding hyperspherical functions can be
related using the ${\cal T}$-coefficients~\cite{tt1,tt2}. 
Schematically, these coefficients relate the following tree structures
\begin{equation}
   \begin{minipage}{0.25\linewidth}
  \includegraphics[width=\linewidth]{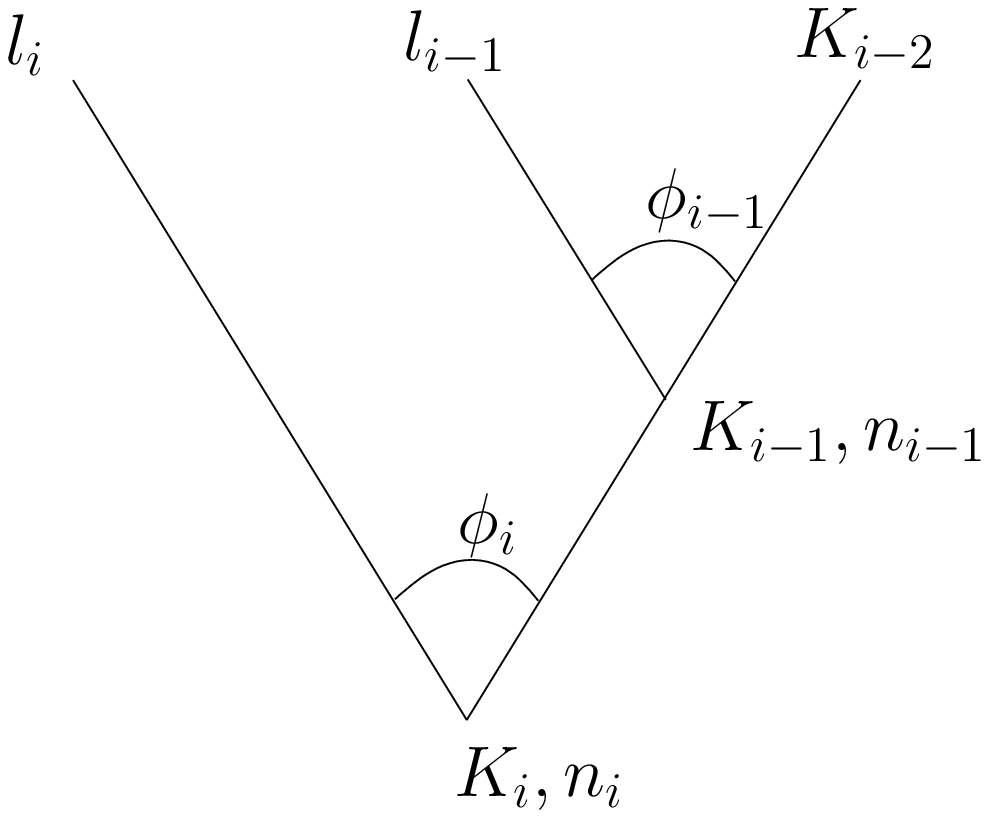}
   \end{minipage}
   =
   \sum_{\tilde n_{i-1}=0}^{N_i}
   {\cal T}^{\alpha_{K_{i-2}}\alpha_{l_{i-1}}\alpha_{l_i}}_{n_{i-1} \tilde n_{i-1}
   K_i}
   \begin{minipage}{0.25\linewidth}
  \includegraphics[width=\linewidth]{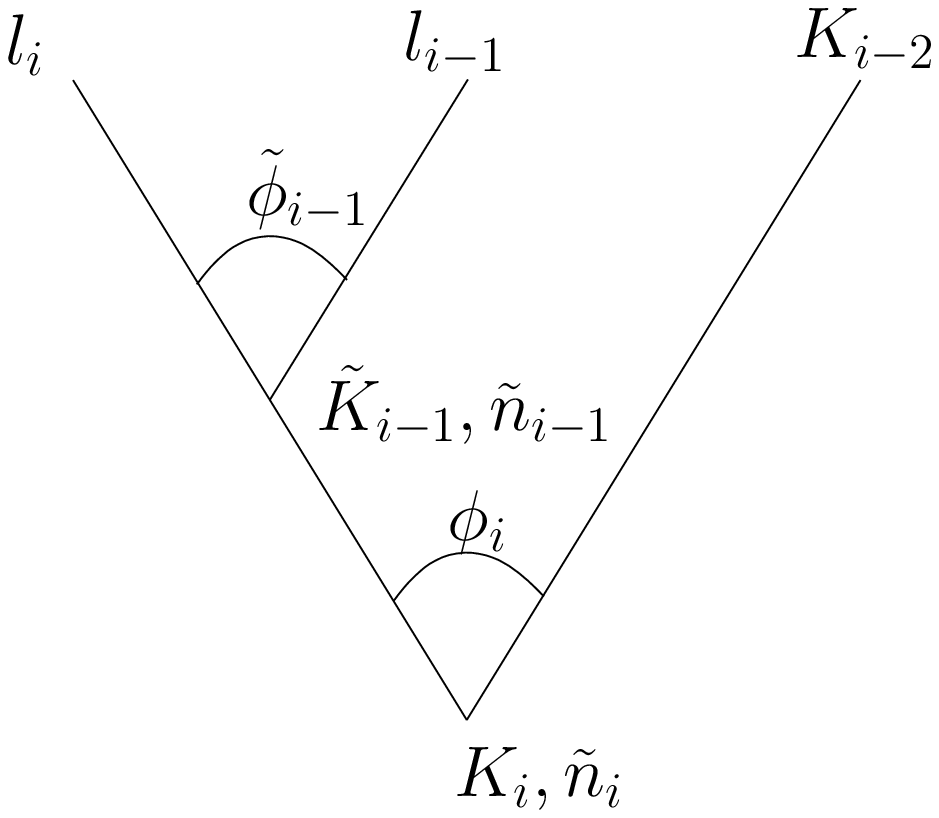} 
   \end{minipage} \,,
\end{equation}
and play the same r\^ole of three-momenta recoupling as the $6j$
coefficients, but for the grand-angular momenta. 
Here $K_i=K_{i-1}+l_i+2n_i=\tilde K_{i-1}+l_i+2\tilde n_i$
The explicit definition of the coefficients is
\begin{equation}
  \begin{aligned}
   {\cal T}^{\alpha_{K_{i-2}}\alpha_{l_{i-1}}\alpha_{l_i}}_{n_{i-1} \tilde n_{i-1}
   K_i}
    =
    & 
    \frac{%
    {\cal N}_{n_{i-1}}^{\alpha_{K_{i-2}} \alpha_{l_{i-1}}}
    \,{\cal N}_{n_{i}}^{\alpha_{K_{i-1}} \alpha_{l_{i}}}
    \begin{pmatrix} n_{i-1} + \alpha_{K_{i-2}} \\ n_{i-1} \end{pmatrix}
      }{%
    {\cal N}_{\tilde n_{i-1}}^{\alpha_{l_{i-1}} \alpha_{l_{i}}}
    \,{\cal N}_{\tilde n_{i}}^{\alpha_{K_{i-2}} \alpha_{\tilde K_{i-1}}}
    \begin{pmatrix} \tilde n_{i}+\alpha_{K_{i-2}} \\ \tilde n_{i} \end{pmatrix} 
      }
      \left(\frac{1}{2}\right)^{n_{i-1}}  \\
      &\frac{2\tilde n_{i-1}+\alpha_{l_{i}}+\alpha_{l_{i-1}}+1}{2^{\alpha_{l_{i}}+\alpha_{l_{i-1}}+1}}
      \frac{\tilde n_{i-1}!\,\Gamma(\tilde n_{i-1}+\alpha_{l_{i}}+\alpha_{l_{i-1}}+1)}
      {\Gamma(\tilde n_{i-1}+\alpha_{l_{i}}+1)\Gamma(\tilde n_{i-1}+\alpha_{l_{i-1}}+1)}
 \\
     & \int_{-1}^{1} dy\,
     (1-y)^{\alpha_{l_{i-1}}+n_{i-1}}(1+y)^{\alpha_{l_{i}}}
    P_{n_{i}}^{\alpha_{K_{i-1}}\alpha_{l_{i}}}\left(y\right)
    P_{\tilde n_{i-1}}^{\alpha_{l_{i-1}}\alpha_{l_{i}}}\left(y\right)\,.
  \end{aligned}
  \label{eq:tree6j}
\end{equation}
In this expression the value of the coefficients $\alpha_j$ depend on the value
of the partial grand angular momentum $K_j$ or partial angular momentum $l_j$, 
which labels the node or the leave respectively, and on the topology of the hyperspherical 
tree. 
Having in mind that in a binary tree a node and its child nodes form a
sub-binary tree with ${\cal N}_{\alpha_j}$  nodes and ${\cal L}_{\alpha_j}$
leaves the coefficients  read
\begin{equation}
  \alpha_j = j + {\cal N}_{\alpha_j} + \frac{1}{2}{\cal L}_{\alpha_j}  \,.
   \label{}
 \end{equation}

Furthermore, the integral in Eq.~(\ref{eq:tree6j})
can be rewritten as an
hypergeometrical function using the following identity:
\begin{equation}
  \begin{aligned}
 & \int_{-1}^1 dy\, (1-y)^\tau(1+y)^\beta P_n^{(\alpha,\beta)}(y)
  P_m^{(\rho,\sigma)}(y) =
  \frac{\Gamma(\alpha-\tau+n)\Gamma(\beta+n+1)\Gamma(\rho+m+1)\Gamma(\tau+1)}
  {\Gamma(\rho+1)\Gamma(\alpha-\tau)\Gamma(\beta+\tau+n+2)}\\
  &\times 
  \frac{2^{\beta+\tau+1}} {m! n!}
  \;\;{}_4F_3\begin{bmatrix} -m, \rho+\sigma+m+1, \tau+1, \tau+1, \tau-\alpha+1 \\
                         \rho+1, \beta+\tau+n+2, \tau-\alpha-n+1
                       \end{bmatrix}\,.
  \end{aligned}
  \label{}
\end{equation}
For the sake of completeness, we also report the notation we use
for the recoupling of three angular momenta
\begin{equation}
   \Big[[Y_{l_{i-2}}({\hat x}_{i-2})
   Y_{l_{i-1}}({\hat x}_{i-1})]_{L_{i-1}}Y_{l_i}({\hat x}_i)\Big]_{L_i}=
   \sum_{\widetilde L_{i-1}} 
    T^{l_{i-2}l_{i-1}l_i}_{L_{i-1} \widetilde L_{i-1}
   L_i}
   \left[Y_{l_{i-2}}({\hat x}_{i-2})
   [Y_{l_{i-1}}({\hat x}_{i-1})Y_{l_i}({\hat x}_i)]_{\widetilde
   L_{i-1}}\right]_{L_i} \,,
   \label{eq:recoupling_l}
\end{equation}
where we have defined 
\begin{equation}
    T^{l_{i-2}l_{i-1}l_i}_{L_{i-1} \tilde L_{i-1} L_i}
    = 
    (-1)^{l_{i-2}+l_{i-1}+l_i+L_i}
    \sqrt{2L_{i-1}+1}\,\sqrt{2\tilde L_{i-1}+1}
  \begin{Bmatrix}
    l_{i-2} & l_{i-1} & L_i \\
    l_i & L_i & \tilde L_{i-1}
  \end{Bmatrix} \,.
\end{equation}

Both $T$-  and ${\cal T}$-coefficients have particular relevance in the construction
of HH functions with arbitrary permutational symmetry~\cite{barnea99, barnea97}.

\subsection{Rotation matrices between HH basis elements of different Jacobi coordinates}

The Jacobi coordinates explicitly depend on the way of
numbering the A particles. In particular, for an equal mass system,
we have selected a successive order in
Eq.~(\ref{eq:jc2}) starting from the definition of
$\mathbf x_{N} = \mathbf r_2 - \mathbf r_1$. In the following we will refer
to this set as the reference Jacobi set. However, different choices are possibles,
starting for example from $\mathbf x_{N} = \mathbf r_j - \mathbf r_i$,
with the related HH functions depending differently on the particle variables.
In general, the Jacobi coordinates can be defined from a permutation 
$\{p\equiv p_1\ldots p_A \}$ of the $A$ particles, 
$\mathbf r_{p_1}\ldots \mathbf r_{p_A}$, 
resulting in a re-definition of the Jacobi coordinates in which  $\mathbf r_i$,
on Eq.~(\ref{eq:jc2}), is changed to $\mathbf r_{p_i}$. The associated HH
functions, ${\mathcal Y}^{LM}_{[K]}(\Omega^p_N)$, are still defined 
by Eq.~(\ref{eq:hh0}). The explicit indication of the index $p$ of the permutation 
allows to trace back the dependence on the particle variables. 
It is a general property of the HH basis that elements constructed using
a permutation $p$ in the arrangement of the particles can be expressed as
a linear combination of HH basis elements defined using some other order, both
having the same grand angular quantum number.
In our case, we use the HH basis constructed with the 
reference Jacobi set to express 
bases constructed with other arrangements. Accordingly, the property reads
\begin{equation}
    {\mathcal Y}^{LM}_{[K]}(\Omega^p_N) =
     \sum_{[K']} C^{p,LM}_{[K][K']} {\mathcal Y}^{LM}_{[K']}(\Omega_N)\,,
\label{eq:cc1}
\end{equation}
where the sum runs over all quantum numbers compatible with the condition
$K=K'$. As indicated, in the transformation the total angular momentum $LM$
is conserved. For a given number of particles, $N_K$ denotes
the number of HH functions having the same value of $K$. 
Consequently,
the coefficients of the transformation $C^{LM}_{[K][K']}$ form a matrix
of dimension $N_K\times N_K$.
For $A=3$ these matrix elements are the Raynal-Revai
coefficients~\cite{rr}, whose expression is explicitly known.
For $A>3$ the coefficients cannot be given in
a close form, and a few methods have been derived for their
calculations~\cite{novo94,krivec90,viv98,efros95}. 

Here we are interested in a particular set of coefficients relating the 
reference HH basis 
to a basis in which the ordering of two adjacent
particles have been 
transposed. It is easy to verify that there are $A-1$ sets of Jacobi
coordinates of this kind based on the following ordering of the particles:
$(\mathbf r_1,\ldots ,\mathbf r_A, \mathbf r_{A-1})$,
$(\mathbf r_1,\ldots ,\mathbf r_{A-1}, \mathbf r_{A-2},\mathbf r_A),\ldots,$
$(\mathbf r_1,\mathbf r_3, \mathbf r_2,\ldots,\mathbf r_A)$,
$(\mathbf r_2,\mathbf r_1,\ldots,\mathbf r_A)$. This last ordering results in a 
Jacobi set in which all the Jacobi vectors are equal to those of the reference 
set except the last one,
$\mathbf x_N$, which is now $\mathbf x'_N= \mathbf r_1- \mathbf r_2$.
The other $A-2$ orderings lead to $N-1$ Jacobi sets that differ, with respect 
to the original Jacobi set, in the definition of two Jacobi vectors. In fact,
given the transposition between particles $j,j+1$, only the Jacobi vectors
$\mathbf x_i$ and $\mathbf x_{i+1}$, with $i=N-j+1$, are different. 
We label them $\mathbf x'_i$ and $\mathbf x'_{i+1}$, and explicitly they are
\begin{equation}
  \begin{aligned}
 \mathbf x'_{i}  &= - \frac{1}{j} \,\mathbf x_i + 
                      \frac{\sqrt{(j+1)^2-2(j+1)}}{j}  \,\mathbf x_{i+1} \\
 \mathbf x'_{i+1}&=   \frac{\sqrt{(j+1)^2-2(j+1)}}{j} \,\mathbf x_i 
                    + \frac{1}{j} \,\mathbf x_{i+1}   \,,
  \end{aligned}
  \label{eq:jc3}
\end{equation}
with $i=1,\ldots,N-1$. The value $i=1$ corresponds to the transposition of the
pair $(\mathbf r_{A-1}, \mathbf r_A)$, whereas the value $i=N-1$ 
corresponds to the transposition of the pair $(\mathbf r_2, \mathbf r_3)$.
Let us call ${\mathcal Y}^{LM}_{[K]}(\Omega^i_N)$ the HH basis element
constructed in terms of a set of Jacobi coordinates in which the $i$-th
and $i+1$-th Jacobi vectors are given from Eq.(~\ref{eq:jc3}) with all the
other vectors equal to the original ones (transposed basis). 
The case $i=N$ corresponds to the special case, mentioned before,
in which all the vectors are equal except $\mathbf x_N$.
The coefficients
\begin{equation}
 {\mathcal A}^{i,LM}_{[K][K']}=\int d\Omega_N[{\mathcal Y}^{LM}_{[K]} (\Omega_N)]^*
 {\mathcal Y}^{LM}_{[K']}(\Omega^i_N)\,,
\label{eq:ca1}
\end{equation}
are the matrix elements of a matrix ${\mathcal A}^{LM}_i$
 that allows to express the transposed HH basis 
elements in terms of the reference basis. They are a particular case of the
general $C^{p,LM}_{[K][K']}$ defined in Eq.(~\ref{eq:cc1}) and, therefore,
the total angular momentum as well as the grand angular quantum number $K$
are conserved in the above integral ($K=K'$). 
The coefficients
${\mathcal A}^{i,LM}_{[K][K']}$ can be calculated analytically using the
$T$- and ${\cal T}$- coupling coefficients
and the Raynal-Revai matrix elements~\cite{novo94,tt1,tt2} .
In fact, we have seen that only two Jacobi coordinates are changed in the 
construction of the transposed HH basis (see Eq.~(\ref{eq:jc3})). If the two
coordinates are directly coupled both in grand-angular and angular space, as
is the case for the pair $\mathbf x_1$ and $\mathbf x_2$, corresponding to $i=1$,
the coefficient reduces to the Raynal-Revai coefficient. Explicitly,
\begin{equation}
 {\mathcal A}^{1,LM}_{[K][K']}=\delta_{K,K'}
 \left[\prod_{i=3}^N\delta_{l_i,l^\prime_i}
 \delta_{L_{i-1},L^\prime_{i-1}}\delta_{K_{i-1},K^\prime_{i-1}}\right]
 {\cal R}^{K_2,L_2}_{l_2l_1,l^\prime_2l^\prime_1} \,,
\label{eq:aa1}
\end{equation}
with 
\begin{eqnarray}
 {\cal R}^{K,L}_{l_2l_1,l^\prime_2l^\prime_1}=
 &\int (\cos\phi\sin\phi)^2 d\phi \;^{(2)}{\cal P}^{l_2,l_1}_K(\phi)
 \int d\hat x_1d\hat x_2 [Y_{l_1}(\hat x_1)\otimes Y_{l_2}(\hat x_2)]^*_{LM} \cr
 & \times \;^{(2)}{\cal P}^{l'_2,l'_1}_K(\phi')
[Y_{l'_1}(\hat x'_1)\otimes Y_{l'_2}(\hat x'_2)]_{LM} \,,
\label{eq:rr1}
\end{eqnarray}
whose analytic form has been given in Ref.~\cite{rr}.
When $2 \le i \le N-1$, we still have a transformation between only two Jacobi 
coordinates, and the coefficients read
\begin{equation}
 {\mathcal A}^{i,LM}_{[K][K']}=
 \left[\prod_{j=1}^{i-1}\delta_{l_j,l^\prime_j}
 \prod_{k=2}^{i-1}\delta_{L_{k},L^\prime_{k}}\delta_{K_{k},K^\prime_{k}}\right]
  \,^{(i)}\!{\cal A}^{L_{i-1}K_{i-1}, L_{i+1} K_{i+1}}_
  {l_i,l'_i,l_{i+1},l'_{i+1},L_i K_i,L'_i K'_i}
 \left[\prod_{j=i+2}^{N}\delta_{l_j,l^\prime_j}
  \prod_{k=i+1}^{N}
 \delta_{L_{k},L^\prime_{k}}\delta_{K_k,K^\prime_k}\right] \,,
\label{eq:aa2}
\end{equation}
where $L_N=L$ and $K_N=K$. From the conservation of partial angular and grand angular momenta
and the fact that $x_i^2+x_{i+1}^2=x'^2_i+x'^2_{i+1}$,
the matrices ${}^{(i)\!}\!{\cal A}$ can be obtained from
a three-dimensional integral. However, 
as has been shown in Ref.~\cite{tt2}, 
they can be reduced to Raynal-Revai coefficients
using the $T$- and ${\cal T}$-coefficients to recouple the
quantum numbers relative to the 
Jacobi variables $\mathbf x_i$ and $\mathbf x_{i+1}$.
The final expression is 
\begin{equation}
  ^{(i)}\!{\cal A}^{L_{i-1}K_{i-1}, L_{i+1} K_{i+1}}_
  {l_i,l'_i,l_{i+1},l'_{i+1},L_i K_i,L'_i K'_i}= 
\sum_{\tilde L_{i}}T^{L_{i-1}l_il_{i+1}}_{L_{i}{\tilde L_{i}}L_{i+1}}\;
T^{L_{i-1}l'_il'_{i+1}}_{L'_{i}{\tilde L_{i}}L_{i+1}}
 \sum_{\tilde n_{i}} 
 {\cal T}^{\alpha_{K_{i-1}}\alpha_{l_{i}}\alpha_{l_{i+1}}}_{n_{i} \tilde n_{i}
  K_{i+1}}
  {\cal T}^{\alpha_{K_{i-1}}\alpha_{l'_{i}}\alpha_{l'_{i+1}}}_{n'_{i} \tilde n_{i}
  K_{i+1}}
 {\cal R}^{\tilde K_{i},\tilde L_{i}}_{l_{i+1}l_i,l'_{i+1}l'_i} \,,
\label{eq:ttr}
\end{equation}
where $\tilde K_{i}= l_i+l_{i+1} + 2\tilde n_{i}$.
Finally, the case $i=N$ corresponds to the transposition of particles $(1,2)$ 
resulting in
$\mathbf x'_N=-\mathbf x_N$ and the coefficient reduces to a simple phase factor
\begin{equation}
 {\mathcal A}^{N,LM}_{[K][K']}=(-1)^{l_N}\delta_{[K],[K']} \,.
\end{equation}

We are now interested in obtaining the rotation coefficients between the
reference HH basis and a basis in which the last
Jacobi vector is defined as $\mathbf x'_N=\mathbf r_j-\mathbf r_i$,
without loosing generality we consider $j>i$. 
A generic rotation coefficient of this kind can be
constructed as successive products of the ${\mathcal A}^{k,LM}_{[K][K']}$
coefficients. For $j\ge 3$, if $\mathbf x_{N-j+2}$ is the last Jacobi vector in 
which
particle $j$ appears, at maximum $2(j-2)$ factors have to be included in the
product. To see this, we start discussing the case $j=3$ resulting in two 
different bases, one having the
vector $\mathbf x'_N=\mathbf r_3-\mathbf r_1$ and the other the vector
$\mathbf x'_N=\mathbf r_3-\mathbf r_2$.
The rotation coefficient that relates the basis having the vector
$\mathbf x'_N=\mathbf r_3-\mathbf r_1$ to the reference basis, is 
${\mathcal A}^{N-1,LM}_{[K][K']}$ since it corresponds to the transposition of
particles $(2,3)$. In the second case we have to consider the transpositions of
particles $(2,3)$ and $(1,2)$ and, the coefficients results the
same as before times the phase given by the coefficient 
${\mathcal A}^{N,LM}_{[K][K']}$. Therefore, in the case $j=3$, the
rotation coefficients includes at maximum the multiplication
of two ${\cal A}$-coefficients. There are three vectors
$\mathbf x'_N=\mathbf r_j-\mathbf r_i$ with $j=4$. When $i=1$ or $i=2$, the transposition
$(3,4)$ leads to the previous case and two and three factors are needed respectively.
For the case $\mathbf x'_N=\mathbf r_4-\mathbf r_3$, the intermediate transposition
$(2,3)$ is needed with the consequence that the rotation coefficient includes
four factors, and so on.

Defining ${\mathcal Y}^{LM}_{[K]}(\Omega^{ij}_N)$ the HH basis element
constructed in terms of a set of Jacobi coordinates in which the 
$N$-th Jacobi vector is defined $\mathbf x'_N=\mathbf r_j-\mathbf r_i$,
the rotation coefficient relating this basis to the reference basis
can be given in the following form
\begin{equation}
 {\mathcal B}^{ij,LM}_{[K][K']}=\int d\Omega[{\mathcal Y}^{LM}_{[K]} (\Omega_N)]^*
 {\mathcal Y}^{LM}_{[K]}(\Omega^{ij}_N) =
\left[{\mathcal A}^{LM}_{i_1}\cdots{\mathcal A}^{LM}_{i_n}\right]_{[K][K']} \,.
\label{eq:ca2}
\end{equation}
The particular values of the indices $i_1,\ldots,i_n$, labelling
the matrices ${\mathcal A}^{LM}_{i_1},\ldots,{\mathcal A}^{LM}_{i_n}$, 
depend on the pair $(i,j)$.
The number of factors cannot be greater than $2(j-2)$ and it increases,
at maximum, by two units from $j$ to $j+1$. 
The matrix
\begin{equation}
{\mathcal B}_{ij}^{LM}={\mathcal A}^{LM}_{i_1}\cdots{\mathcal A}^{LM}_{i_n}\,,
\label{eq:matrixb}
\end{equation}
is written as a product of the sparse matrices ${\mathcal A}^{LM}_{i}$'s, 
a property which 
is particularly well suited for a numerical implementation of the
potential energy matrix as is discussed in the next section.

\section{The potential energy matrix in terms of the ${\cal A}$-coefficients}\label{sec:pot}

The potential energy of an $A$-body system constructed in terms of two-body
interactions reads
\begin{equation}
   V=\sum_{i<j} V(i,j)  \;\;\; .
\end{equation}
Considering the case of a central two-body interaction, its matrix 
elements in terms of the HH basis of Eq.(\ref{eq:hh0}) are
\begin{equation}
   V_{[K][K']}(\rho)=\sum_{i<j} 
\langle{\cal Y}^{LM}_{[K]}(\Omega_N)|V(i,j)|{\cal
Y}^{LM}_{[K']}(\Omega_N)\rangle \, .
\end{equation}
In each element $\langle{\cal Y}^{LM}_{[K]}|V(i,j)|{\cal Y}^{LM}_{[K']}\rangle$ the integral
is understood on all the hyperangular variables and depends parametrically on 
$\rho$. Explicitly, for the pair $(1,2)$, it results
\begin{equation}
\begin{aligned}
& V^{(1,2)}_{[K][K']}(\rho)=
\langle{\cal Y}^{LM}_{[K]}(\Omega_N)|V(1,2)|{\cal Y}^{LM}_{[K']}(\Omega_N)\rangle= \cr
&\delta_{l_1,l^\prime_1}\cdots\delta_{l_N,l^\prime_N}
\delta_{L_2,L^\prime_2}\cdots\delta_{L_N,L^\prime_N}
\delta_{K_2,K^\prime_2}\cdots\delta_{K_N,K^\prime_N} \cr
&\times \int d\phi_N(\cos\phi_N\sin\phi_N)^2
\;{}^{(N)}{\cal P}^{l_N,K_{N-1}}_{K_N}(\phi_N)
V(\rho\cos\phi_N)\;{}^{(N)}{\cal P}^{l_N,K_{N-1}}_{K'_N}(\phi_N)\,.
\end{aligned}
\label{eq:v12}
\end{equation}
The above formula shows that for $A>2$ the matrix representation
of $V(1,2)$ is sparse in this basis. Using the rotation coefficients,
a general term of the potential $V(i,j)$ results
\begin{equation}
\begin{aligned}
 V^{(i,j)}_{[K][K']}(\rho)=
\langle{\cal Y}^{LM}_{[K]}(\Omega_N)|V(i,j)|{\cal
Y}^{LM}_{[K']}(\Omega_N)\rangle=  \cr
\sum_{[K''][K''']}{\cal B}^{ij,LM}_{[K''][K]}{\cal B}^{ij,LM}_{[K'''][K']}
\langle{\cal Y}^{LM}_{[K'']}(\Omega^{ij}_N)|V(i,j)|{\cal
Y}^{LM}_{[K''']}(\Omega^{ij}_N)\rangle \,.
\end{aligned}
\label{eq:vij}
\end{equation}
It should be noticed that
\begin{equation}
\langle{\cal Y}^{LM}_{[K]}(\Omega_N)|V(1,2)|{\cal
Y}^{LM}_{[K']}(\Omega_N)\rangle= 
\langle{\cal Y}^{LM}_{[K]}(\Omega^{ij}_N)|V(i,j)|{\cal
Y}^{LM}_{[K']}(\Omega^{ij}_N)\rangle \,,
\end{equation}
therefore Eq.(\ref{eq:vij}) results
\begin{equation}
\begin{aligned}
 V^{(i,j)}_{[K][K']}(\rho)=
\langle{\cal Y}^{LM}_{[K]}(\Omega_N)|V(i,j)|{\cal
Y}^{LM}_{[K']}(\Omega_N)\rangle=  \cr
\sum_{[K''][K''']}{\cal B}^{ij,LM}_{[K''][K]}{\cal B}^{ij,LM}_{[K'''][K']} 
 V^{(1,2)}_{[K''][K''']}(\rho) \,,
\end{aligned}
\label{eq:vijp}
\end{equation}
or, in matrix notation,
\begin{equation}
  V_{ij}(\rho)= [{\cal B}^{LM}_{ij}]^{t} \,V_{12}(\rho)\,{\cal B}^{LM}_{ij}\,.
\label{eq:mij}
\end{equation}
The complete potential matrix energy results
\begin{equation}
\sum_{ij} V_{ij}(\rho)=\sum_{ij} 
[{\cal B}^{LM}_{ij}]^t\, V_{12}(\rho)\,{\cal B}^{LM}_{ij} \,.
\label{eq:vpot}
\end{equation}
The matrices ${\cal B}^{LM}_{ij}$ are block matrices with each block
labelled by the grand angular momentum $K$. Moreover,
each block is constructed as a product of the sparse matrices 
${\mathcal A}^{LM}_{i}$ as defined in Eq.(\ref{eq:matrixb}). On the other
hand the matrix $V_{12}(\rho)$, defined in Eq.(\ref{eq:v12}), couples 
different values of $K$ but it is diagonal in the quantum numbers related to
particles $3,\ldots, A$. 

Each term of the sum in Eq.(\ref{eq:vpot}) results in a product of sparse
matrices, a property which allows an efficient implementation of matrix-vector
product, key ingredient in the solution of the Schr\"odinger equation using
iterative methods.

\section{Results for $A=3-6$ systems }\label{sec:results}

In this section we present results for $A=3-6$ systems obtained by
a direct diagonalization of the Hamiltonian of the system. The
corresponding Hamiltonian matrix is obtained using 
the following orthonormal basis
\begin{equation}
  \langle\rho\,\Omega\,|\,m\,[K]\rangle =
  \bigg(\beta^{(\alpha+1)/2}\sqrt{\frac{m!}{(\alpha+m)!}}\,
  L^{(\alpha)}_m(\beta\rho)
  \,{\text e}^{-\beta\rho/2}\bigg)
  {\cal Y}^{LM}_{[K]}(\Omega_N)  \,,
  \label{mhbasis}
\end{equation}
where $L^{(\alpha)}_m(\beta\rho)$ is a Laguerre polynomial with
$\alpha=3N-1$ and $\beta$ a variational non-linear parameter.
The matrix elements of the Hamiltonian are obtained after
integrations in the $\rho,\Omega$ spaces. They depend on 
the indices $m,m'$ and $[K],[K']$ as follows
\begin{equation}
\begin{aligned}
  \langle m'\,[K']|H|\,m\,[K] \rangle = -\frac{\hbar^2\beta^2}{m}
 ( T^{(1)}_{m'm}-K(K+3N-2) T^{(2)}_{m'm}) \delta_{[K'][K]} \cr
 + \sum_{ij} \left[
\sum_{[K''][K''']}{\cal B}^{ij,LM}_{[K][K'']}{\cal B}^{ij,LM}_{[K'''][K']}
 V^{m,m'}_{[K''][K''']}\right] \,.
\end{aligned}
\label{eq:hmm}
\end{equation}
The matrices $T^{(1)}$ and $T^{(2)}$ have an analytical form
and are given in Ref.~\cite{mario}.  The matrix elements
$V^{m,m'}_{[K][K']}$ are obtained after integrating the matrix $V_{12}(\rho)$
in $\rho$-space (we will call the corresponding matrix $V_{12}$).
Introducing the diagonal matrix $D$ such that
$\langle [K']\,|\,D\, | [K]\rangle = \delta_{[K],[K']} K(K+3N-2)$, and the identity
matrix $I$ in $K$-space, we can rewrite the Hamiltonian schematically as
\begin{equation}
  H = -\frac{\hbar^2\beta^2}{m} ({}^{(1)}T \otimes I  +  {}^{(2)}T\otimes D )
  + \sum_{ij} [{\cal B}^{LM}_{ij}]^t\, V_{12}\,{\cal B}^{LM}_{ij} \,,
  \label{eq:schemH}
\end{equation}
in which the tensor product character of the kinetic energy is explicitly
given. A scheme to diagonalize such a matrix  is given in the
Appendix.

We choose as central potential
the Volkov potential
\begin{equation}
 V(r)=V_R \,{\rm e}^{-r^2/R^2_1} + V_A\, {\rm e}^{-r^2/R^2_2} \,,
\end{equation}
with $V_R=144.86$ MeV, $R_1=0.82$ fm, $V_A=-83.34$ MeV, and $R_2=1.6$ fm. 
The nucleons are considered to have the same mass chosen to be equal to the
reference mass $m$ and corresponding to
$\hbar^2/m = 41.47~\text{MeV\,fm}^{-2}$.
With this parametrization of the potential, the
two-nucleon system has a binding energy $E_{2N}=0.54592\;$MeV and a
scattering length $a_{2N}=10.082\;$fm.
This potential has been used several times in the literature making its
use very useful to compare different methods
\cite{barnea99,varga95,viviani05,timo02}. The use of central
potentials in general produces too much binding, in particular the
$A=5$ system results bounded. Conversely, the use of the 
$s$-wave version of the potential produces a spectrum much closer
to the experimental situation. This is a direct consequence of the
weakness of the nuclear interaction in $p$-waves. Accordingly,
we analyze both versions of the potential, the central
Volkov potential and the $s$-wave projected potential.
The results are obtained
after a direct diagonalization of the Hamiltonian matrix of
Eq.(\ref{eq:hmm}) including $m_{max}+1$ Laguerre polynomials with a fix
value of $\beta$, and all
HH states corresponding to maximum value of the grand angular momentum
$K_{max}$. The scale parameter $\beta$ can be used as a non-linear
parameter to study the convergence in the index $m=0,1,\ldots,m_{max}$, with
$m_{max}$ the maximum value considered. In
the present analysis the convergence will be studied with respect to
the index $K_{max}$, therefore, the number of Laguerre polynomials at
each step, $m_{max}+1$, will be sufficiently
large to guarantee independence from $\beta$ of the physical eigenvalues
and eigenvectors. We found that $m_{max}+1\approx 20$ Laguerre polynomials (with proper
values of $\beta$) were sufficient for an accuracy of $0.1$\% 
in the calculated eigenvalues.

\subsection{Symmetries of the eigenvectors}

Fixing the total angular momentum and parity $J^\pi$ of the state we want to
describe, the diagonalization of the Hamiltonian produces eigenvectors with
well-defined-permutation symmetry. Since we are using a central potential, the
total angular momentum $L$ and total spin $S$ are good quantum numbers.
Accordingly, our basis is identified by $(L,S,T)J^\pi$, where $T$ is the total
isospin of the state, and the parity corresponds to consider even or odd $K$
values in the expansion.  The eigenvalues appear either in singlets,
corresponding to symmetric or antisymmetric eigenvectors, or in multiplets,
corresponding to mixed symmetry eigenvectors. The identification of the symmetry
of each eigenvector can be done applying to it the Casimir operator
\begin{equation}
C(A)=\sum_{i<j} P(i,j) \,,
\label{eq:casimir1}
\end{equation}
where $P(i,j)$ is the permutation operator of particles $(i,j)$. Using the results
of the preceding section, the representation of the Casimir operator in
the HH basis results
\begin{equation}
C(A)=\sum_{i<j} {\cal B}^{LM}_{ij} (-1)^{L_N} B^{LM}_{ij}\,.
\label{eq:casimir2}
\end{equation}
As discussed in Ref.~\cite{novo95}, this Casimir operator corresponds to the class
sum $[(2)]_A$ of the group of permutation of $A$ objects
$S_A$ and the corresponding eigenvalues $\lambda$ for the different 
symmetries $[{\bm \lambda}]$ are given in that reference up to $A=5$.
The eigenvectors of the Hamiltonian are also eigenvectors of this Casimir
operator, therefore the application of the Casimir operator to a specific
eigenvector $\Psi^{L^\pi}_n([{\bm \lambda}])$ results
\begin{equation}
 C(A)\Psi^{L^\pi}_n([{\bm \lambda}])=\lambda \Psi^{L^\pi}_n([{\bm \lambda}]) \,.
\end{equation}
The different symmetries characterizing the spatial eigenvector
are identified by $\lambda$. The physical state of $A$ nucleons is
obtained after multiplying $\Psi^{L^\pi}_n([{\bm \lambda}])$
by the proper spin-isospin state in order to obtain an antisymmetric state.

\subsection{$A=3,4$ systems}

In Ref.~\cite{mario} the binding energies $E_0(3)$ and $E_0(4)$
corresponding to the ground states of the $A=3,4$ systems has been
studied using the Volkov potential. 
Here we extend the analysis to some more
states of the spectrum using, in addition,
the $s$-wave version of the potential. 
In particular the $A=3$ system present a very shallow excited state.
This is a consequence of the very shallow two-nucleon binding energy
$E_{2N}$ and the large value of the scattering lenth $a_{2N}$ that 
this potential produces. It is known that when the two-body system 
presents these characteristics, the
three-body system could show a certain numbers of bound states close
to the two-body threshold called Efimov states (see Ref.~\cite{hammer}
and reference therein). In the present case, this behavior is a
consequence of the parametrization of the Volkov potential that has been 
tuned to approximate the binding
energy of the $A=3$ system. In doing that, the binding energy
of the $A=2$ system results to be much lower than the experimental deuteron
binding energy. Despite this unrealistic situation,
here we are interested in studying the
HH expansion for systems with $A>4$. The analysis of the $A=3,4$
systems serves as a basis for establishing the different thresholds
that appear in the description of those systems. 

In Table~\ref{tab:A3T1} the $A=3$ results
for the state $(L,S,T)J^\pi=(0,1/2,1/2)1/2^+$ are given using the
complete potential as well as its $s$-wave version. The ground state binding energy
$E_0$ converges at the level of $0.1$ keV with $K_{max}=40$ and, fixing
the non-linear parameter $\beta=2\;{\rm fm}^{-1}$, with $m_{max}=24$.
For the sake of comparison the results of the stochastic variational model (SVM)
of Ref.~\cite{varga95} and those 
from Ref.~\cite{barnea99} are given in the table.
The convergence of the binding energy $E_1$ of the shallow state at the same level of accuracy
necessitates a much larger basis. The maximum grand angular quantum
number has been increased up to $K_{max}=320$ and, with 
$\beta=1\;{\rm fm}^{-1}$, the maximum degree of Laguerre polynomials
used was $m_{max}=32$. Above $K_{max}=60$ only symmetric states with $l_1=l_2=0$
have been considered. This very different pattern of convergence in
the two binding energies, $E_0$ and $E_1$, has
been observed before~\cite{barletta09,prl09}. Moreover,
the pattern of convergence of the all-waves
and $s$-wave potentials is similar. Since the structure of these
states corresponds mostly to have the particles in a relative
$l=0$ state, there is only a small decrease in energy for $E_0$,
of about $35$ keV,
when the $s$-wave potential is considered. In the excited state
$E_1$ this difference is even less, of about $0.5$ keV,
giving both versions of the potential very close values. 
This is a manifestation of the particular structure 
of the Efimov state in which the third particle orbitates around
the $l=0$ state of the other two in a very far orbit.
Increasing the attraction of the two-body potential, the two-body
binding energy $E_{2N}$ increases faster than the $E_1$ energy
and, at some point, the Efimov state starts to be above the
two-body threshold (see for example Ref.~\cite{barletta01}). 
When realistic forces are used to describe the three-nucleon system
there is no observation of an excited state, in agreement with
the experimental situation. However the effective range function 
presents a pole close to the two-body threshold~\cite{kievsky97}, 
that can be interpreted as an Efimov-like state embedded in the continuum.
When the Coulomb interaction is included, the ground state binding energy results
$E_0=7.7594$ MeV (all-waves potential) and $E_0=7.7254$ MeV
($s$-wave potential). In
both cases the isospin components $T=1/2$ and $3/2$ are automatically included. 
With the repulsion induced by the Coulomb potential the excited state 
is not any more bounded.

The $L=0$ state of the $A=4$ system is firstly analyzed. The
spatially symmetric state of four nucleons can be antisymmetrized
using the $S=0,T=0$ spin-isospin functions.
In Table~\ref{tab:A4T1} the pattern of convergence, in terms of $K_{max}$,
is shown for the first two levels of the $(L,S,T)J^\pi=(0,0,0)0^+$ state
using both versions of the Volkov potential. 
The ground state binding energy $E_0$ converges at
the level of $1-2$~keV for $K_{max}=40$ whereas the convergence of the
excited state binding energy $E_1$ has been estimated at the level of $50$~keV. 
For both types of potentials the excited state results to be bounded with respect
to the $3+1$ threshold.
For the sake of comparison the results of the SVM
and those from Ref.~\cite{barnea99} are shown in the table.
In order to compare the results to the experimental value of the
$\alpha$-particle, $B({}^4{\rm He})=28.30$~MeV, the last four columns of
the table show the results including the Coulomb potential between the two
protons. The obtained values of $29.60$~MeV (all-waves
potential) and $29.43$~MeV ($s$-wave potential) show a pronounced
overbinding.
This is the usual situation when central interactions are used to 
describe the $^4$He nucleus and it is at variance to the case in which
realistic NN forces are used.
When the Coulomb potential is included the excited state
appears slightly above the $3+1$ threshold. In the case of the all-waves 
potential the lowest threshold, corresponding to a $p$-$^3$H configuration,
is at $8.465$~MeV whereas for the $s$-wave potential it results to be
at $8.431$~MeV. The $n$-$^3$He thresholds are at $7.759$~MeV and
$7.725$~MeV respectively. Though the convergence
for $E_1$ was not completely achieved, the description is
close to the experimental observation of a $0^+$ resonance between
both thresholds and centered 395 keV above the $p$-$^3$H threshold. 

Let us consider the negative parity $L=1$ state. The lowest level
corresponds to the $[\bm{3\;1}]$ irreducible representation and can be antisymmetrized using
the $S=1,T=0$ or $S=0,T=1$ spin-isospin functions of four nucleons. 
Accordingly, using a
central potential, the $J^\pi=0^-,1^-,2^-$ states are degenerated.
The results are given in Table~\ref{tab:A4T2}. We can observe
that the all-wave potential produces a bound state
at approximate $10.4$~MeV far from the experimental observation
of a $0^-$ resonance $800$~keV above the $0^+$ resonance. 
Conversely, using the $s$-wave potential the level results to be unbounded.
It appears at approximate $1.4$~MeV above the 3+1 threshold and at
approximate $1.3$~MeV above the $0^+$ resonance in better agreement with the
experimental situation. When the Coulomb potential between the two protons 
is considered the triple degeneracy of the $[\bm{3\;1}]$ representation
breaks in three different levels, $E_0$, $E_1$, $E_2$, showed
in the last three columns of the table. The state corresponding to the $E_0$ level 
is formed by an
antisymmetric proton pair times a symmetric neutron pair and can be
completely antisymmetrized with the spin state of the two protons having $S_p=1$ and
the spin state of the two neutrons having $S_n=0$, having total spin $S=1$.
The state corresponding to the $E_1$ level is formed by a
symmetric proton pair times a symmetric neutron pair and can be
completely antisymmetrized with the spin of the two protons $S_p=0$ and
the spin of the two neutrons $S_n=0$, having total spin $S=0$. Finally,
the state corresponding to the $E_2$ level is formed by a
symmetric proton pair times an antisymmetric neutron pair and can be
completely antisymmetrized with the spin of the two protons $S_p=0$ and
the spin of the two neutrons $S_n=1$, having total spin $S=1$. 
The first and third level are mostly $T=0$ and can be identified with the
$(J^\pi,T)=(0^-,0)$ and $(2^-,0)$ resonances whereas the $E_1$ level is mostly $T=1$ and
can be identify with the $(1^-,1)$ resonance, observed in the low
energy spectrum of $^4$He~\cite{tilley92}.

\begin{figure}
  \begin{center}
 \includegraphics[width=\linewidth]{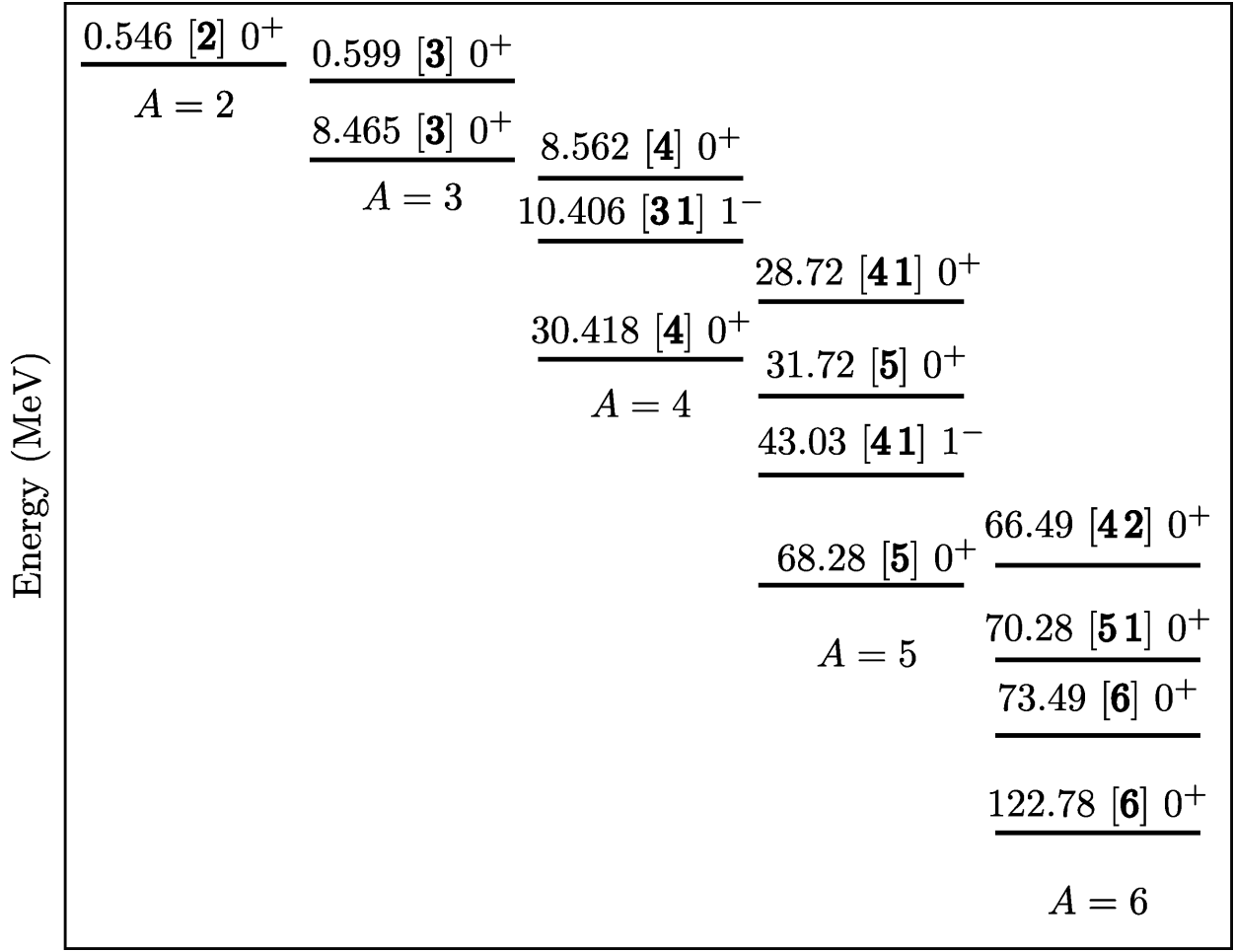}
  \end{center}
  \caption{Calculated levels for $A=2-6$ using the all-waves Volkov potential.}
 \label{fig:fig2}
\end{figure}

\begin{figure}
  \begin{center}
 \includegraphics[width=\linewidth]{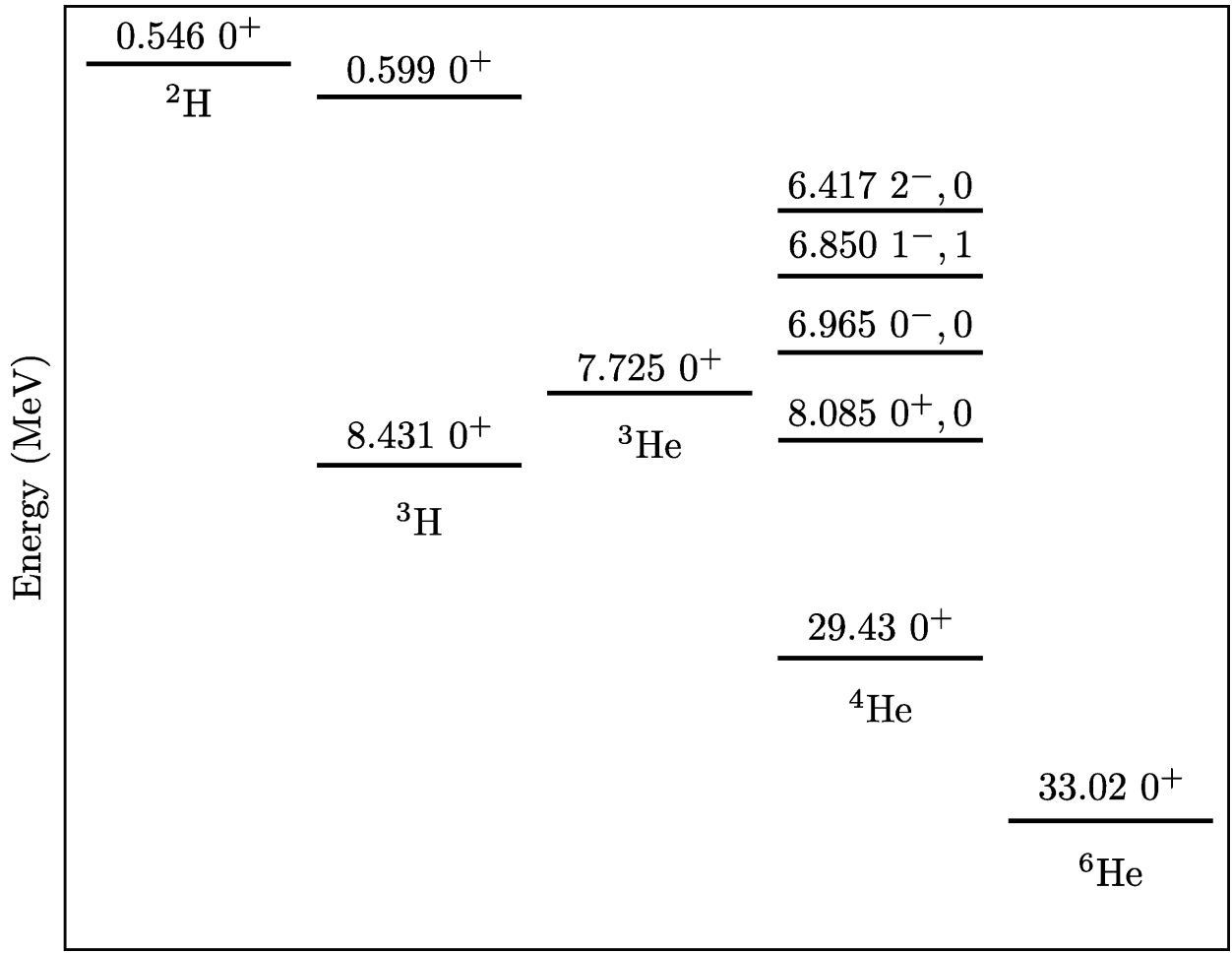}
  \end{center}
  \caption{Calculated levels for $A=2,3,4$ and 6, using the $s$-wave Volkov
  potential with the inclusion of Coulomb interaction for He isotopes.
  In this case the $A=5$ system results unbounded.}
 \label{fig:fig3}
\end{figure}

We can conclude that besides its simplicity, the $s$-wave potential
describes the $A=4$ system better than the complete potential and,
in some cases, in reasonable agreement with the experiment.
From a technical point of view we were able to describe the $L=0$ ground and
first excited states and the first level of the $L=1$ state
using the non-symmetrized HH functions. In particular the
convergence of the $L=0$ first excited state, $E_1$, presents some difficulties
since its energy results to be very close to the threshold. 
When the Coulomb interaction is taken into account this level moves to the
continuum and it appears as a resonance between the two $3+1$
thresholds, in agreement with the experimental data. The $s$-wave
potential describes better also the negative parity resonances; moreover,
in order to accurately extract their position and width, the present method can
be combined with the procedure developed for example in Ref.~\cite{witala99}.
The computed states for $A=3,4$ are collected in
Figs.~\ref{fig:fig2},\ref{fig:fig3}.
 
\subsection{$A=5,6$ systems}

\begin{figure}
  \begin{center}
 \includegraphics[width=\linewidth]{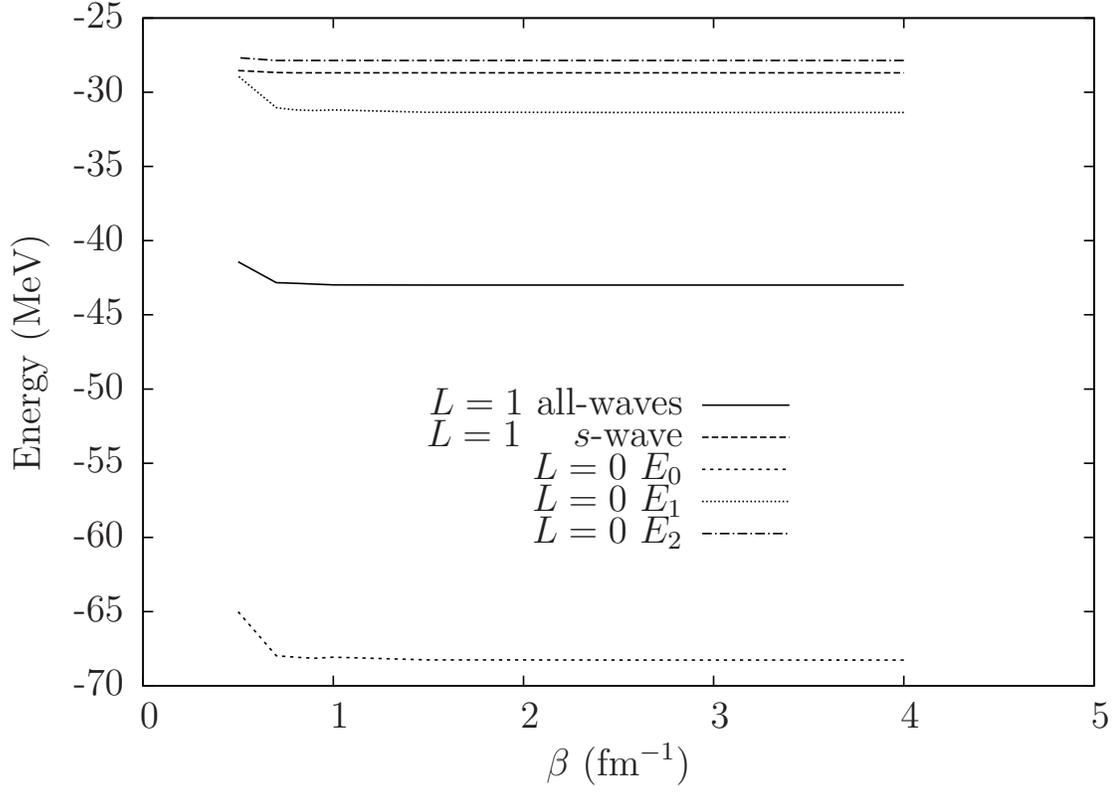}
  \end{center}
  \caption{The $A=5$, $L=0$ levels, given in Table~\protect\ref{tab:A5T1},
denoted as $E_0,E_1,E_2$ and the $L=1$ levels given in Table~\protect\ref{tab:A5T2},
denoted as all-waves and s-wave, are shown as functions of the non-linear parameter $\beta$,
at $K_{max}=16$ ($L=0$ levels) and $K_{max}=17$ ($L=1$ levels), respectively.}
 \label{fig:fig4}
\end{figure}

In the case of systems with $A>4$ the spatially-symmetric state cannot
be antisymmetrized using the corresponding spin-isospin functions. Therefore,
it is interesting to study the symmetry of the different levels in the 
$A=5$ system when the non-symmetrized basis is used. For the positive parity
$L=0$ state, using the Volkov potential, we found that the deepest
two levels correspond to a completely symmetric state (the irreducible
representation of $S_5$ $[\bm{5}]$) as expected. As mentioned, they cannot be
antisymmetrized using the spin-isospin functions of five particles
and, therefore, they do not represent physical states for five nucleons. 
The third level belongs to the irreducible representation of $S_5$
$[\bm{4\;1}]$; it can be antisymmetrized using the $A=5$ spin-isospin
functions having $S=1/2,T=1/2$, and accordingly it represents the lowest level
of the $(L,S,T)J^\pi=(0,1/2,1/2)1/2^+$ state of five nucleons. 
The convergence of
these three states in terms of $K_{max}$ is given in Table~\ref{tab:A5T1}.
The first two levels, representing bosonic bound states,
present a good convergence with $K$ (in particular the deepest level).
The convergence of the $[\bm{4\;1}]$ state shows that it does not
describe a bound state, in agreement with the fact
that the $A=5$ nucleus does not exist. In fact 
its energy results to be above the threshold of $30.42$~MeV 
describing an $^4$He nucleus plus a fifth nucleon far away (here the
Coulomb interaction has not been included). 
For the three levels, their
stability as a function of the non-linear parameter $\beta$ is shown in
Fig.~\ref{fig:fig4}.

The negative parity $L=1$ state corresponds to the
$(L,S,T)J^\pi=(1,1/2,1/2)1/2^-$ and $(1,1/2,1/2)3/2^-$ states which
are degenerate using the Volkov potential. Its deepest level
cannot be spatially-symmetric, and in fact it belongs to the $[\bm{4\;1}]$
representation; as before, it can be antisymmetrized using the 
$S=1/2,T=1/2$ spin-isospin functions of five nucleons. 
In Table~\ref{tab:A5T2} the convergence
of this level is shown in terms of $K_{max}$ for the all-waves
potential as well as its $s$-wave reduction. From the table we
can observe that the all-waves version of the potential 
predicts a very deep bound state, at $43.03$~MeV, whereas
the $s$-wave reduction does not. Using the $s$-wave potential, the
$A=5$ systems results to be unbounded 
in agreement with the experimental observation. From this analysis
we can conclude that the fact that the $A=5$ nucleus does not exist is the
result of a delicate balance between the Pauli principle, the
short range character of the NN interaction and its weakness
in $p$-waves. The $s$-wave potential, used in the present analysis,
represents the extreme case in which
the interaction in $p$-waves is considered zero.
For these two levels, their
stability as a function of the non-linear parameter $\beta$ is shown in
Fig.~\ref{fig:fig4} as the solid line (all-waves) and long dashed line
(s-wave) respectively.
In the last three columns of the table, the three
levels obtained considering the $s$-wave Volkov potential plus the
Coulomb interaction between two protons are shown. The inclusion of the
Coulomb interaction breaks the degeneracy of the quartet-$[\bm{4\;1}]$
state, producing three different states that can be identified
by the residual $S_2\otimes S_3$ symmetry of the 
two-protons and three-neutrons sub-systems.
The lowest two states $E_0$ and $E_1$ belong to the $[\bm{1}^2]\otimes[\bm{3}]$ 
and $[\bm{2}]\otimes[\bm{3}]$ representations of $S_2 \otimes S_3$
and cannot be antisymmetrized with respect to the three neutrons. 
The third level, $E_2$, is a doublet
state since corresponds to a mixed symmetry of the three neutrons 
and it is symmetric in the two protons. Its belongs to the
$[\bm{2}]\otimes[\bm{2\; 1}]$ representation and
it can be antisymmetrized with
the spin state of the two protons having $S_p=0$ and the spin state of the three neutrons
having $S_n=1/2$. Physically this state is describing a scattering state between
a neutron and an $\alpha$-particle in $J^\pi=1/2^-$ and $3/2^-$.
In the present study we are limiting the description to
bound states, however, using the method described in Ref.~\cite{kievsky10}
it would be possible to compute phase-shifts using the $L=0,1$
bound-like states. 
The extension of the method to describe scattering states is in progress.

In the case of the $A=6$ system we concentrate the analysis in the 
$(L,S,T)J^\pi=(0,0,1)0^+$ and $(0,1,0)1^+$ states. Using a
central potential, and disregarding the Coulomb interaction, 
these two states are degenerate. Including the Coulomb interaction 
between two protons, the first state has the quantum numbers of $^6$He. 
A direct diagonalization of the six body Hamiltonian using
the non-symmetrized HH basis, with the Volkov potential, produces a spectrum in which
the first two levels belongs to the $[\bm{6}]$ irreducible representation
of $S_6$. They are completely symmetric and cannot be antisymmetrized
using the $A=6$ spin-isospin functions. The third level
belongs to the $[\bm{5\;1}]$ representation
and it cannot be antisymetrized too. 
The fourth level belongs to the $[\bm{4\;2}]$ representation,
and it is the first one that can be symmetrized using the $A=6$ spin-isospin 
functions having $S=0,T=1$ or $S=1,T=0$.
The convergence pattern of these four levels in terms of $K_{max}$
are shown in Table~\ref{tab:A6T1} indicated by $E_i$, $i=1,\ldots,4$.
Similar to the $A=5$ case, the Volkov potential acting in all waves
predicts large binding energies. In particular the binding energy of the
physical state results to be $\approx 67$~MeV. 
Using the $s$-wave potential a much more
reasonable value of $\approx 34$~MeV is obtained for this level. The
corresponding convergence is shown in the last column
of Table~\ref{tab:A6T1} indicated by $E^s_3$. It should be noticed
that in the computation of the spectrum using the $s$-wave potential
the $E^s_3$ is not anymore the fourth level. Other levels belonging
to the $[\bm{6}]$ and $[\bm{5\;1}]$ representation gain
more energy than the $[\bm{4\;2}]$ level, making difficult its
correct identification. However, it is possible to restrict the
search of the eigenvectors to those having a particular
symmetry using a symmetry-adapted Lanczos method~\cite{slanczos} 
(a description of the iterative method used is given in the Appendix).
Essentially, starting with a vector having the desired symmetry, 
after each iteration of the matrix-vector product, the new vector is
projected onto the sub-space of the selected symmetry. 
Following Ref.~\cite{slanczos}, an intermediate
purification step is also implemented. This method has the
characteristic of finding eigenvalues corresponding to
eigenvectors of one particular symmetry simplifying the search
procedure and the identification of the eigenvectors.

When the Coulomb interaction between two protons is considered
the degeneracy of the $[\bm{4\;2}]$ level (of dimension 9)
is broken and four different states appear. It is possible
to identify the physical state looking at the symmetry of the
four neutrons. One of the states belongs to the $[\bm{4}]$
representation, two belong to the $[\bm{3\;1}]$ representation
and the last one belongs to the $[\bm{2}^2]$ representation of $S_4$.
This last state is the only one that can be antisymmetrized using
the spin functions of four neutrons having $S_n=0$. Moreover, the
proton state is spatially symmetric and therefore can be antisymmetrized
with the spin function $S_p=0$ making a total $S=0$ state. The convergence
of this state is given in the last column of Table~\ref{tab:A6T1}.
It should be noticed that this state is embedded in a very dense
spectrum. In order to follow this state in the projected Lanczos
method a projection-purification procedure is performed. Essentially
the vector, after each matrix vector product, 
is projected on antisymmetric states between particles (3,4) and (5,6).
In this way the level belonging to the $[\bm{2}]\otimes[\bm{2}^2]$
representation of $S_2\otimes S_4$ results to be the lowest one.
The results obtained for the different levels are collected in
Figs.~\ref{fig:fig2},\ref{fig:fig3}.

\section{Conclusions}\label{sec:conclu}

In this work we have developed a technique devoted to describe bound
states in an $A$-body system without imposing a particular requirement
due to the intrinsic statistic of the particles. However, the final aim of
the method is to found wave functions that fulfill this requirement.

Starting with the non-symmetrized HH basis set, we have diagonalized the
Hamiltonian of the $A$-body system using that basis at fixed values of $K$.  We
have observed that the eigenvectors reflect the symmetries present in the
Hamiltonian and, in particular, if the system is composed by identical
particles, the eigenvectors belong to the different irreducible representations
of the permutation group of $A$ objects, $S_A$. Using a Casimir operator, it
was possible to identify those eigenvectors having the required symmetry of the
system and, accordingly, study the convergence (in terms of $K$) of the
corresponding eigenvalues. The direct use of the non-symmetrized HH basis has
important consequences from a technical point of view. The size of the basis is
much bigger than the one limited to a subspace having a particular symmetry.
However, it should be noticed that a system of nucleons includes spatial, spin
and isospin degrees of freedom,  all of them coupled by the NN potential, with the
consequence that different spatial symmetries are present in an $A$-nucleon wave
function. Although the construction of HH basis elements having different
spatial symmetries is possible (see for example Ref.~\cite{barnea99}), the
necessity of including the different symmetries in the description enlarges the
dimension of the basis and makes it comparable to the case in which the
non-symmetrized basis is used. This is particularly important when one wants to
consider the description of the small components of the wave function induced by
symmetry breaking terms in the potential, as for example high isospin
components. 

The method here presented is based in a particular implementation of
the potential energy matrix constructed as a sum of products of sparse
matrices. This allows to efficiently use iterative algorithms in which 
the matrix-vector product is a key element. However the iterative
methods are well suited to calculate the deepest levels of the
spectrum. In our formulation, due to the presence of different symmetries, the
physical states could appear very high in the spectrum or in a zone with a
high density of levels. In this case
we found very convenient to use the symmetry-adapted Lanczos method~\cite{slanczos}. 
Using the particular form of the permutation operator $P(i,j)$ in terms of
the sparse matrices (see Eq.(\ref{eq:casimir2})), it was possible to
project the vector in the iterative procedure to be antisymmetric 
in selected pairs of particles. In this way the desired symmetry becomes
the lowest state of the spectrum. Though this mechanism is not as fast
as searching for the true lowest state of the complete spectrum, it is
much faster than searching for certain numbers of levels in high
position of the spectrum.

We should also stress that the sparse matrices
$^{(i)}{\cal A}$ defined in Eq.(\ref{eq:ttr}) have the 
property of being constructed as products of the angular $T$-coefficients,
the tree ${\cal T}$-coefficients and the Raynal-Revai coefficients.
Whereas the latters couples quantum number belonging to the $[K]$ and $[K']$ 
sets, the $T$- and ${\cal T}$-coefficients perform a recoupling
of quantum numbers inside $[K]$ or $[K']$. Moreover the
Raynal-Revai coefficients ${\cal R}^{K,L}_{l_2l_1,l^\prime_2l^\prime_1}$
couple quantum numbers belonging to three particles
(see Eq.(\ref{eq:rr1})). As the number of particles $A$ increases, 
more values of the quantum numbers $K,L$ are accessible and this makes
the size of the $^{(i)}{\cal A}$ matrices to
increase, though slowly with the number of particles $A$. Furthermore,
going from a system with $A$ particles to $A+1$, the number of potential
terms increase by $A$ and the number of factors in the matrix ${\cal B}_{ij}^{LM}$ 
of Eq.(\ref{eq:matrixb}) increases at maximum
of the same quantity. The computational effort increases roughly linear
with $A$ and this fact makes feasible the application of
the method for increasing values of $A$ as has been demonstrated in the
present work. Our expectation is that the present technique could be
extended to treat systems up to $A=8$. The calculations presented here have been
obtained using a sequential code.
We expect that an opportune parallelization of the code (which is under study)
will increase the potentiality of the method. 

We have limited the analysis to consider a central potential, the Volkov
potential, used several times in the literature. 
Though the use of a central potential leads to an unrealistic description of the
light nuclei structures, the study has served to analyze the
characteristic of the method: the capability of the diagonalization
procedure to construct the proper symmetry of the state and the particular structure, 
in terms of products of sparse matrices, of the Hamiltonian matrix. The
success of this study makes feasible
the extension of the method to treat interactions depending on spin and isospin 
degrees of freedom as the realistic NN potentials. A preliminary analysis
in this direction has been done~\cite{mario09}.
To this respect it is important
to notice that the information of the potential is given in the
matrix $V_{12}$ defined in Eq.(\ref{eq:v12}). Once this matrix is
given, the method remains the same, with the basis enlarged to include
spin and isospin degrees of freedom if required. Using the Volkov
potential we have shown that it was possible to identify all the
physical states and the corresponding thresholds in order to interpret 
the level as bounded or belonging to the continuum. Furthermore,
the results obtained using the Volkov potential up to $A=6$ compare well with 
other techniques. A few characteristic of the $A=3-6$ systems using
the Volkov potential are the following.
Due to its particular parametrization a shallow state appears
in the $A=3,4$ systems when the Coulomb interaction is not considered.
In the $A=3$ this state has the characteristic of an Efimov state.
When the Coulomb interaction is considered these states move
to the continuum. The Volkov potential acting in all waves
produces large binding energies
as $A$ increases. Accordingly we have included in the analysis the $s$-wave
version of the potential. In agreement with the experimental observations,
this version predicts in reasonable positions
the $A=4$ $0^+$ and $0^-$ resonances and no bound states in the $A=5$ system. 
It also predicts reasonable binding energies in the $A=6$ system.
The extension of the method to consider realistic potentials is in progress.

\section{Appendix}
The diagonalization of the Hamiltonian is obtained by means of an iterative
algorithm which requires only the action of the Hamiltonian matrix
on a given vector. We used the Lanczos algorithm in the version invented by Cullum and
Willoughby~\cite{cullum} 
which is particularly sparing with memory use. 
In principle, the iterative procedure should preserve the permutation symmetry
of the input vector, as the Hamiltonian commutes with the group elements. 
However, the round-off errors generate components also
in the other irreducible representations.
To circumvent this problem, we have used
a symmetry adapted Lanczos (SAL) developed in Ref.~\cite{slanczos}
in which a projection operator is applied after each iterative step.
Starting from a random initial vector, in the usual
Lancsoz recurrence formula
\begin{equation}
\beta_{i+1}{\bm v}_{i+1}=H{\bm v}_i
 -\alpha_i{\bm v}_i-\beta_i{\bm v}_{i-1} \,,
\end{equation}
the product $H{\bm v}_i$ is replaced by $P^{[{\widehat{\bm \lambda}}]}H{\bm
v}_i$,
where $P^{[\widehat{\bm \lambda}]}$ is a projector on a sub-space with a
non-zero intersection with  the irreducible representation $[{\bm \lambda}]$,
and zero intersection with the irreducible representations of the
lower-eigenvector symmetries.
A purification step is also performed in which the product
$\beta_{i+1}{\bm v}_{i+1}$ is replaced by 
$P^{[\widehat{\bm \lambda}]}\beta_{i+1}{\bm v}_{i+1}$.

As an example in the $L=0$ sector of the $A=6$ system, we are interested in states
belonging to the irreducible representation $[{\bm 4\,\bm 2}]$. In order to
eliminate lower states belonging to the irreducible representations $[{\bm 6}]$
and $[{\bm 5\,\bm 1}]$, we have used the projector 
\begin{equation}
  P^{\widehat{[{\bm 4\;\bm 2}]}}  = A_{12} \cdot A_{34}\,,
  \label{}
\end{equation}
given as the product of the antisymmetrization operator $A_{12}$ with respect
particles $(1,2)$, and the antisymmetrization operator 
$A_{34}$ with respect particles $(3,4)$.
The two antisymmetrization operators have the following expression
in terms of the ${\cal A}_i$ matrices (the superscript $L,M=0,0$ is understood)
\begin{equation}
  A_{12} = \frac{1}{2}(1-{\cal A}_5),
  \label{}
\end{equation}
and
\begin{equation}
  A_{34} = {\cal A}_4 {\cal A}_3 {\cal A}_4 \, 
  \frac{1-{\cal A}_5}{2}
   \, {\cal
  A}_4 {\cal A}_3 {\cal A}_4 \,. 
  \label{}
\end{equation}

\newpage

\begin{table}[h]
  \caption{$A=3$ results for
$(L,S,T)J^\pi=(0,\frac{1}{2},\frac{1}{2})\frac{1}{2}^+$ state
   using the all-waves and $s$-wave Volkov potential as a function 
   of the maximum grand angular quantum number $K_{max}$.
   The ground state $E_0$ as well as the
   excited state $E_1$ are shown.}
  \label{tab:A3T1}
\begin{center}
  \begin{tabular}{l c c| c c}
  \hline
            &\multicolumn{2}{c|}{all-waves} & \multicolumn{2}{c}{$s$-wave} \\
  $K_{max}$ &  $E_0$ (MeV) & $E_1$ (MeV)& $E_0$ (MeV) & $E_1$ (MeV) \\
  \hline
    20      &  8.4623      &  0.3627   &   8.4283     & 0.3618    \\
    40      &  8.4649      &  0.5181   &   8.4309     & 0.5174    \\
    60      &  8.4649      &  0.5595   &   8.4309     & 0.5589    \\
    80      &  8.4649      &  0.5773   &   8.4309     & 0.5768    \\
   100      &              &  0.5866   &              & 0.5861    \\
   120      &              &  0.5918   &              & 0.5913    \\
   140      &              &  0.5947   &              & 0.5943    \\
   160      &              &  0.5965   &              & 0.5960    \\
   180      &              &  0.5976   &              & 0.5971    \\
   200      &              &  0.5982   &              & 0.5978    \\
   240      &              &  0.5989   &              & 0.5985    \\
   280      &              &  0.5992   &              & 0.5988    \\
   320      &              &  0.5993   &              & 0.5989    \\
  \hline
   SVM~\protect\cite{varga95}  & 8.46  &           &    &           \\
   Ref.~\protect\cite{barnea99}  & 8.462 &   0.2599     &  &     \\
  \hline
\end{tabular}
\end{center}
\end{table}

\begin{table}[h]
  \caption{Binding energies for the $A=4$ ground state $E_0$ and the first excited 
   state $E_1$ of the $(L,S,T)J^\pi=(0,0,0)0^+$ state
   using the all-waves and $s$-wave Volkov potentials as a function 
   of the maximum grand angular quantum number $K_{max}$. In the last
   four columns the Coulomb interaction has been considered. For the sake of
   comparison the results of
   Refs.~\protect\cite{barnea99,varga95} are shown.}
  \label{tab:A4T1}
\begin{center}
  \begin{tabular}{c c c c c |c c c c}
  \hline
   &\multicolumn{2}{c}{all-waves} & \multicolumn{2}{c|}{$s$-wave}
  &   \multicolumn{2}{c}{all-waves} & \multicolumn{2}{c}{$s$-wave} \\
  $K_{max}$ &  $E_0$ (MeV) & $E_1$ (MeV)& $E_0$ (MeV) & $E_1$ (MeV) &
  $E_0$ (MeV) & $E_1$ (MeV)& $E_0$ (MeV) & $E_1$ (MeV) \\
  \hline
   0  & 28.580 & 3.238 & 28.580 & 3.238 & 27.748 & 2.787 & 27.748 & 2.787 \\
   10 & 30.278 & 7.509 & 30.116 & 7.445 & 29.456 & 7.039 & 29.292 & 6.976 \\
   20 & 30.416 & 8.223 & 30.250 & 8.164 & 29.596 & 7.778 & 29.429 & 7.720 \\
   30 & 30.418 & 8.463 & 30.252 & 8.403 & 29.599 & 8.035 & 29.431 & 7.976 \\
   40 & 30.418 & 8.562 & 30.252 & 8.501 & 29.600 & 8.144 & 29.432 & 8.085 \\
   \hline
   SVM~\protect\cite{varga95}& 30.42  &   &       & & &  &  &       \\      
   Ref.~\protect\cite{barnea99}  & 30.406 & 8.036 & & &  & & &  \\
   \hline
\end{tabular}
\end{center}
\end{table}

\begin{table}[h]
  \caption{The binding energy of the $A=4$ lowest level having $L=1$,
   using the all-waves and $s$-wave Volkov potential, as a function
   of the maximum grand angular quantum number $K_{max}$.
   In the case of $s$-wave potential, when Coulomb interaction between particles
   (1,2) is considered, the level splits in three sub-levels,
   whose energies 
   $E_0$, $E_1$ and $E_2$
   are shown in the last three columns.}
  \label{tab:A4T2}
\begin{center}
  \begin{tabular}{c c c|c c c}
  \hline
  $K_{max}$ &  all-waves & $s$-wave & $E_0$ (MeV) & $E_1$ (MeV) & $E_2$ (MeV) \\
  \hline
   1 & 7.965 & 0.387 &- & -  & -  \\
   3 & 8.411 & 1.975 & 1.639 & 1.440 & 1.374 \\
   11& 10.121& 5.567 & 5.314 & 5.091 & 4.899 \\
   21& 10.373& 6.642 & 6.456 & 6.276 & 5.955 \\
   31& 10.406& 7.113 & 6.965 & 6.850 & 6.417  \\
   \hline
   \hline
\end{tabular}
\end{center}
\end{table}

\begin{table}[H]
  \caption{$A=5$ binding energies of the first three levels of the
  $L=0$ state, belonging to the indicated irreducible representation
  $[\bm{\lambda}]$, as a function of $K_{max}$.
  The size of the HH basis $N_{HH}$ is also indicated.}
  \label{tab:A5T1}
\begin{center}
  \begin{tabular}{c c c c c}
  \hline
  \hline
  $K_{max}$ \rule{0pt}{12pt} &  $N_{HH}$   & $E_0$ (MeV)&
  $E_1$ (MeV)& $E_2$ (MeV)  \\
   &     & [$\bm{5}$] & [$\bm{5}$] & [$\bm{4\;1}$]  \\
  \hline \\
   0 & 1     & 64.864 & 24.472 &  -\\
   2 & 10    & 64.864 & 24.472 & 20.160 \\
   4 & 55    & 65.958 & 28.411 & 22.043 \\
   6 & 220   & 66.893 & 29.517 & 24.415 \\
   8 & 714   & 67.713 & 30.228 & 25.568 \\
   10 & 1992 & 68.008 & 30.587 & 26.459 \\
   12 & 4950 & 68.177 & 30.927 & 27.043 \\
   14 & 11220 & 68.239 & 31.152& 27.515 \\
   16 & 23595 & 68.264 & 31.357& 27.862 \\
   18 & 46618 & 68.274 & 31.509& 28.143 \\
   20 & 87373 & 68.278 & 31.628 & 28.371 \\
   22 & 156520 & 68.279& 31.715 & 28.560 \\
   24 & 269620 & 68.280& 31.779 & 28.719 \\
   \hline
\end{tabular}
\end{center}
\end{table}

\begin{table}[H]
  \caption{$A=5$ binding energies of the deepest $L=1$ state,
  as a function of $K_{max}$,
  using the all-waves and $s$-wave Volkov potential. In the last three
columns the Coulomb potential has been summed to the $s$-wave Volkov potential.
  The size of the HH basis $N_{HH}$ is also indicated.}
  \label{tab:A5T2}
\begin{center}
  \begin{tabular}{c c c c | c c c}
  \hline
   $K_{max}$ & $N_{HH}$  & all-waves & $s$-wave  & $E_0$ & $E_1$  & $E_2$ \\
  \hline
   1 & 4     & 39.635  & 21.874 & 21.370 & 21.119 &  -  \\
   3 & 40    &  40.001 & 24.317 & 23.854 & 23.604 & 23.524 \\
   5 & 220   & 41.022  & 26.053 & 25.618 & 25.367 & 25.251 \\
   7 & 876   &  41.785 & 26.923 & 26.505 & 26.258 & 26.116 \\
   9 & 2820  & 42.384  & 27.546 & 27.140 & 26.896 & 26.736 \\
   11 & 7788 & 42.682  & 27.971 & 27.574 & 27.333 & 27.160 \\
   13 &19140 & 42.868  & 28.297 & 27.908 & 27.669 & 27.485 \\
   15 &42900 & 42.952  & 28.521 & 28.140 & 27.903 & 27.710 \\
   17 &89232 & 42.996  & 28.693 & 28.320 & 28.084 & 27.882 \\
   19 &174460 & 43.017 & 28.823 & 28.457 & 28.223 & 28.011 \\
   21 &323752 & 43.027 & 28.924 & 28.562 & 28.331 & 28.110  \\
   23 &574600 & 43.032 & 29.005 & 28.647 & 28.417 & 28.189 \\
  \hline
   SVM&       & 43.00  &        & & &  \\
   HH~\protect\cite{barnea99}  & & 42.383 &  & & &  \\
   \hline
\end{tabular}
\end{center}
\end{table}

\begin{table}[H]
  \caption{$A=6$ binding energies of the first four levels of the
  $L=0$ state, using the Volkov potential,
  belonging to the indicated irreducible representation
  $[\bm{\lambda}]$, as a function of $K_{max}$. $E^s_3$ indicates the
  binding energy of the 
  lowest [$\bm{4\;2}$] state using the $s$-wave Volkov potential, 
  and $E_3^{sc}$ is the binding energy of the $[\bm{2}]\otimes[\bm{2^2}]$ state,
  once the Coulomb interaction has been included.
  The size of the HH basis $N_{HH}$ is also indicated.}
  \label{tab:A6T1}
\begin{center}
  \begin{tabular}{c c c c c c c c}
  \hline
  \hline
  $K_{max}$ \rule{0pt}{12pt} &  $N_{HH}$   & $E_0$ (MeV)&
  $E_1$ (MeV)& $E_2$ (MeV) & $E_3$ (MeV) & $E_3^s$ (MeV) & $E_3^{sc}$ (MeV) \\
   &     & [$\bm{6}$] & [$\bm{6}$] & [$\bm{5\;1}$] & [$\bm{4\;2}$] & [$\bm{4\;2}$] 
   & $[\bm{2}]\otimes[\bm{2^2}]$\\
  \hline \\
   0 &     1       & 117.205& 64.701 & - & - & - & -                      \\
   2 &     15      & 117.205& 64.701 & 62.513 & 61.142 & 24.793 & 24.064  \\
   4 &     120     & 118.861& 69.450 & 64.277 & 62.015 & 28.791 & 28.016  \\
   6 &     680     & 120.345& 70.544 & 66.268 & 63.377 & 30.723 & 29.935  \\
   8 &     3045    & 121.738& 71.443 & 67.280 & 64.437 & 31.645 & 30.851  \\
   10 &   11427    & 122.317& 71.923 & 68.371 & 65.354 & 32.244 & 31.446  \\
   12 &   37310    & 122.597& 72.477 & 69.029 & 65.886 & 32.708 & 31.908  \\
   14 &   108810   &122.711& 72.822 & 69.531 & 66.201 & 33.075 & 32.275  \\
   16 &   288990   &122.752& 73.101 & 69.842 & 66.360 & 33.358 & 32.558  \\
   18 &   709410   &122.768& 73.284 & 70.051 & 66.437 & 33.561 & 32.762  \\
   20 &   1628328  &122.774& 73.407 & 70.189 & 66.474 & 33.710 & 32.912  \\
   22 &   3527160  &122.776 & 73.485 & 70.283 & 66.491& 33.814 & 33.017  \\
   \hline
   SVM&        &       &        & & 66.25 &        \\
   \hline
\end{tabular}
\end{center}
\end{table}

\newpage 
\clearpage

\end{document}